\documentclass[apj,twocolumn]{openjournal}

\usepackage{latexsym}
\usepackage{graphicx}
\usepackage{amssymb}
\usepackage{amsmath}
\usepackage{longtable}
\usepackage{textcomp}
\usepackage{booktabs}
\usepackage{natbib}
\usepackage{hyperref}
\hypersetup{colorlinks=true,linkcolor=blue,citecolor=blue,filecolor=blue,urlcolor=blue}
\usepackage{orcidlink}

\graphicspath{{figures/}}
\providecommand{\degree}{\ensuremath{^{\circ}}}
\newcommand{\altt}{\mathrm{alt}}
\newcommand{\az}{\mathrm{az}}
\newcommand{\vect}[1]{\boldsymbol{#1}}
\newcommand{\Rot}{\mathsf{R}}
\newcommand{\ascal}{\texttt{ascal}}
\newcommand{\px}{\,\mathrm{px}}

\makeatletter
\long\def\@makecaption#1#2{%
 \noindent\begin{minipage}{0.9999\linewidth}%
   \if\csname ftype@\@captype\endcsname 2
   \vskip 2ex\noindent\centering\@table@type@size{\@eapj@cap@font #1}\par
   {\raggedright\footnotesize #2\par}\medskip
   \else
   \vspace*{\abovecaptionskip}\noindent\footnotesize #1 #2\par\vskip \belowcaptionskip
   \fi
 \end{minipage}\par}
\makeatother

\begin{document}

\title{Automatic astrometric calibration of low-cost all-sky cameras}
\shorttitle{Astrometric calibration of low-cost all-sky cameras}

\author{J.~R. Gonz\'alez Fern\'andez\,\orcidlink{0000-0002-4825-8188}$^{1,\star}$}
\author{Laura Hermosa Mu\~noz\,\orcidlink{0000-0002-9610-0123}$^{2}$}
\author{Mar\'ia Fern\'andez Alonso$^{3}$}
\author{Luc\'ia Gonz\'alez Cuesta\,\orcidlink{0000-0002-1241-5508}$^{3}$}

\affiliation{$^1$Universidad Internacional de La Rioja (UNIR), Avenida de la Paz, 137, 26006 Logro\~no, La Rioja, Spain}
\affiliation{$^2$Departamento de F\'{i}sica, Universidad de Oviedo, Campus de Llamaquique, C/ Calvo Sotelo s/n, 33007 Oviedo, Spain}
\affiliation{$^3$Allande Stars, Allande, Asturias, Spain}
\thanks{$^{\star}$E-mail: juanrodrigo.gonzalezfernandez@unir.net}

\shortauthors{Gonz\'alez Fern\'andez et al.}

\begin{abstract}
All-sky cameras support weather monitoring, amateur astronomy, fireball and space-debris detection, and measurements of light pollution. These applications require reliable astrometric calibration despite imperfect camera alignment and variable observing conditions. We present an open-source procedure that combines a compact fisheye model, an exact rotation to account for camera tilt, and automatic star matching. It can initialise a calibration from a single frame given the observing site coordinates and observation time. We evaluate the procedure on nine cloud-free, moonless frames, obtained with a Raspberry~Pi camera and a 1.55~mm lens. Fitting on eight frames and evaluating on the ninth in turn gives a median residual of $0.70\px$ (95\% interval 0.64--0.80~px) across 4172 matches, reduced to $0.60\px$ by two optional decentering parameters. All nine single-frame calibrations pass the quality criteria and predict positions on other frames with a median residual of $0.77\px$. We also apply the clear-sky calibration to 73 frames with varying lunar illumination or cloud cover, obtaining comparable residuals for accepted matches on moonlit frames ($0.62\px$). These results support reusing a calibration while the camera geometry remains unchanged and enough stars remain visible for reliable matching. To test the generality of the single-frame procedure, we calibrated one night frame from each of 23 different all-sky systems, including fireball-network cameras, observatory sky monitors and amateur stations, with images of 0.3 to 77~megapixels in JPEG, FITS and camera-raw formats. All 23 are calibrated from the site and time alone, with a median fit residual of $0.42\px$. Applied to a second frame of the same camera without refitting, the calibrations give a median residual of $0.66\px$ over 18 systems. The procedure is available as the Python package \ascal.
\end{abstract}


\maketitle

\twocolumngrid

\section{Introduction}
\label{sec:intro}

A fisheye camera pointed at the zenith records the whole sky in one frame, and for that reason all-sky cameras have become a standard instrument in several branches of observational astronomy. Fireball networks have used them since the photographic era to triangulate meteoroid trajectories \citep{ceplecha1987,borovicka1995}, and their modern descendants, FRIPON \citep{colas2020fripon}, the Desert Fireball Network \citep{howie2017dfn}, CAMS \citep{jenniskens2011cams} and the Global Meteor Network \citep[GMN;][]{vida2021gmn}, operate hundreds of cameras, many of them built from consumer parts and hosted by amateurs. Observatories use all-sky imagers to monitor cloud cover and sky transparency \citep{long2006,yin2025ali,rui2026lenghusky}; light-pollution studies use them to map the night-sky brightness and its sources \citep{duriscoe2007,jechow2017}; and aeronomy uses them to geo-reference airglow and auroral emission \citep{kapali2025autocal}. In every case the scientific product depends on an astrometric calibration, which is the mapping between pixel coordinates and horizontal coordinates on the sky.

Blind astrometric solvers such as Astrometry.net \citep{lang2010astrometry} are widely used for narrow and intermediate fields. A hemispheric fisheye image additionally requires a projection that represents strong radial distortion across the field, and camera orientation and lens geometry must be estimated together. All-sky plate models have long treated the optical-axis orientation as an exact rotation \citep{borovicka1995}, and the fireball networks have refined the radial and non-symmetric distortion terms of that model \citep{jeanne2019fripon,devillepoix2020gfo}. Among other works, \citet{barghini2019} improved their parametrisation and implemented automatic source identification and catalogue matching for PRISMA cameras, starting from an approximate calibration. Another calibration model is ORION \citep{antuna2022orion}, which provides automatic or manual star identification in image sequences and fits the zenith position, the azimuth offset and a radial polynomial. However, its model assumes a well-levelled camera, so the tilt of the optical axis enters only through the position of the zenith pixel, and its automatic mode starts from a previous calibration or approximate initial parameters. Other approaches address specific observing systems, including the Ali Observatory camera \citep{yin2025ali} and automated star-track calibration for aeronomy imagers \citep{kapali2025autocal}. The Global Meteor Network calibrates its cameras with SkyFit2, part of the open RMS software \citep{vida2021gmn,rms_software}, which fits radial and polynomial distortion models to stars identified on the image, starting from a manually adjusted initial plate. These methods establish both the geometry and the feasibility of automation; they differ in their initialisation requirements, in the complexity of the model, in whether a tilt of a few degrees is represented, and in how the result is validated.

For an unattended monitoring station, calibration must also be practical after installation or maintenance. A camera assembled from consumer parts is usually installed by its host, levelled by eye rather than with instruments, moved while its dome is cleaned, and operated without an observer available to identify stars. Its routine archive may contain only compressed 8-bit images, and most of its frames are taken with the Moon in the sky or with some cloud. A procedure to obtain useful data from an all-sky camera installed under these amateur conditions should recover the camera orientation automatically, use a compact model, quantify how well a calibration predicts positions on subsequent frames, and state what it does when the sky is not clear. A small number of parameters helps constrain the fit, but predictive accuracy must be established by validation rather than inferred from parameter count.

We combine an eight-parameter camera model with automatic catalogue matching and single-frame initialisation. The model comprises two tilt angles, a rotation about the optical axis, the two coordinates of the optical centre, a focal scale and two radial coefficients from the model of \citet{kannala2006}. An optional two-parameter Brown--Conrady term \citep{conrady1919,brown1966} represents decentering distortion. We match DAOStarFinder detections \citep{stetson1987} to Hipparcos catalogue stars \citep{esa1997hipparcos}, after transforming their coordinates to the observation date, using position and brightness to identify likely counterparts. Catalogue positions are corrected for atmospheric refraction. A blind pose search supplies the initial solution, and a quality gate rejects poorly constrained fits. The contribution is the integration of these established components into an open calibration workflow, together with a comparison of tilt treatments, validation across frames and a test on 23 all-sky systems of very different design. The behaviour with the Moon and with clouds is characterised as a secondary result. The implementation is released as the project-independent Python package \ascal{} \citep{ascal_repo}.

We distinguish two complementary calibration workflows. \emph{Single-frame calibration} is intended for rapid, one-shot initialisation on the first day of operation: one image, the site coordinates and the observation time suffice, the image disc is detected automatically, and an eight-parameter base model is fitted. \emph{Multi-frame calibration} is intended for more precise calibration once several suitable images are available, whether from one night or several. It uses a dedicated sky mask, photometric matching filters and either the base model or its ten-parameter extension. On this installation the typical prediction residuals are about 0.8~px for single-frame initialisation and 0.6--0.7~px for multi-frame calibration; Sections~\ref{sec:accuracy} and \ref{sec:single} define the corresponding evaluation statistics. These workflows address different levels of available input and refinement; their accuracy difference cannot be assigned to any one ingredient in isolation.

The main application of this calibration methodology is Lumaria, a light-pollution and sky-monitoring network promoted by Allande Stars in rural Asturias, northern Spain \citep{allandestars_web,lumaria_web}. We evaluate the procedure on its first station, at the Zreizeda Remote Observatory in Cereceda, Allande \citep[ZRO;][]{zro_web}, using a Raspberry~Pi camera with a low-cost M12 fisheye lens installed by its host. We estimate the accuracy using nine cloud-free, moonless frames. We fit the model on eight frames and evaluate its predictions on the ninth, repeating this procedure so that each frame serves once as the test frame. We also use 73 frames from the same archive, taken on different nights under a range of lunar altitudes, illuminated fractions and cloud conditions, to test whether the clear-sky calibration applies to these conditions and whether we can obtain a calibration directly from such frames. Finally, we apply the single-frame procedure to frames from 23 other all-sky cameras, taken from public archives and data releases, to establish how far it generalises beyond the installation on which it was developed.

The paper is organised as follows. Section~\ref{sec:data} describes the instrument and the two data sets. Section~\ref{sec:method} presents the camera model, the detection and matching of stars, the parameter estimation, the single-frame procedure and the validation protocol. Section~\ref{sec:results} gives the fitted parameters and the accuracy on clear test frames, compares the treatments of the tilt, examines the role of the magnitude limit and of the decentering term, and reports the single-frame calibrations. Section~\ref{sec:moon} deals with the moonlit and cloudy frames, and Section~\ref{sec:cameras} with the other all-sky cameras. Section~\ref{sec:discussion} places the results in the context of existing methods, including the astrometric fit of SkyFit2, and discusses the limits of the procedure. Finally, Section~\ref{sec:summary} presents the summary and main conclusions of this work.

\section{Instrument and data}
\label{sec:data}

\subsection{Station and camera}

The station is the first Lumaria prototype, at 43.259\degree{}N, 6.603\degree{}W and 650~m above sea level. The camera is a Raspberry~Pi High Quality Camera \citep{rpihqcamera}, built around the Sony IMX477 back-illuminated CMOS sensor ($4056 \times 3040$ pixels of 1.55~$\mu$m, 1/2.3-inch format) and controlled by a Raspberry~Pi~4. This module is used in home-built and amateur all-sky cameras, for which open-source acquisition software exists \citep{jacquin_allsky}. The lens is an EDATEC ED-LENS-M12-230155-12 M12 fisheye \citep{edatec_lens} with a nominal effective focal length of 1.55~mm, aperture F2.0 and a specified field of view of 195\degree{} on a 1/2.3-inch sensor; the image circle extends slightly beyond the horizon (Fig.~\ref{fig:geometry}A). The camera and its electronics are housed in a weatherproof box with an acrylic dome, fastened to a post with a clamp and levelled by eye, as an amateur host would install it; the residual tilt measured below results from that installation by eye. At night the station captures a frame every minute with an exposure of 20~s at ISO~1600 (analogue gain 16) and uploads 8-bit JPEG images to cloud storage; one frame per hour is used here. Lens, camera module and computer together cost less than 200~euros. For orientation, the fitted radial scale decreases from approximately 17.5~px per degree near the optical axis to 12.1~px per degree at a camera-frame zenith distance of 90\degree{} (Section~\ref{sec:parameters}). Thus one pixel corresponds to approximately 3.4--5.0~arcmin in the radial direction. The scale depends on position and direction on the sensor, so the angular-to-pixel conversions below are approximate reference values.

\subsection{Clear-sky data set}
\label{sec:clearset}

We selected data taken between 1 July and 15 August 2026, one capture per hour between 22:00 and 05:00 local time (CEST, UTC+2). To reduce contamination by lunar glow, we required a lunar altitude $< -5\degree$ relative to the horizon. We also required an automatic cloud fraction estimate of at most 0.07, from the project's night-time cloud analysis based on the visibility of reference stars. We then inspected every candidate visually and removed one frame with visible cloud arcs. The result is nine frames, listed separately by local date and time in Table~\ref{tab:dataset}. Between 5768 and 8934 sources are detected per frame (median 6392). Three frames are distributed with the software as examples: one of this set (8 July, 01:00) and two of the second set described below (9 August, 03:00, with the Moon 0.5\degree{} below the horizon, and 28 July, 01:00, with the Moon at 20\degree{} of altitude and 97\% illuminated).

\begin{table}
\centering
\caption{Clear-sky data set. The Moon column gives its altitude at the time of each frame. Test pairs are catalogue-star and detection matches evaluated using a model fitted on the other eight frames. These stars have magnitude $\le 5.5$ and altitude $\ge 3\degree$.}
\label{tab:dataset}
\footnotesize
\setlength{\tabcolsep}{3pt}
\begin{tabular}{lccr}
\toprule
Date (2026) & Local time & Moon (\degree) & Test pairs \\
\midrule
5 July & 00:00 & $-8.6$ & 462 \\
6 July & 00:00 & $-12.1$ & 487 \\
8 July & 00:00 & $-17.8$ & 475 \\
8 July & 01:01 & $-8.1$ & 516 \\
9 August & 00:00 & $-17.8$ & 440 \\
11 August & 01:00 & $-23.3$ & 494 \\
11 August & 02:00 & $-22.0$ & 440 \\
12 August & 00:00 & $-22.8$ & 422 \\
12 August & 01:00 & $-26.7$ & 436 \\
\midrule Total (9 frames) & & & 4172 \\
\bottomrule
\end{tabular}
\end{table}

\subsection{Moonlit and cloudy frames}
\label{sec:moonset}

The second set contains 73 hourly captures from the same selection interval that do not meet the clear-sky criteria. They span 3 July to 12 August and include frames with the Moon above the horizon (47, altitudes up to 37\degree{} and illuminated fractions from 18\% to 100\%), with the Moon within 5\degree{} below the horizon (7) or with clouds while the Moon was down (19).
We classify the frames according to their estimated cloud fractions into mutually exclusive classes: clear (fraction $\le 0.1$; 32 frames), low cloud coverage ($0.1$--$0.3$; 14), partial ($0.3$--$0.6$; 14), heavy ($0.6$--$0.9$; 3) and overcast ($> 0.9$; 10). These labels refer to the fraction of the sky reported as cloudy, not to the optical thickness of the clouds. We exclude these frames from the reference fits, which combine the nine clear, moonless images. In Section~\ref{sec:moon}, we use them to evaluate predictions from those fits and, separately, calibration from a single frame.

\section{Method}
\label{sec:method}

We first describe the camera model that maps sky directions to pixel coordinates. We then explain how we detect stars, calculate their catalogue positions and match them to image detections to estimate the model parameters. Finally, we describe how to initialise a calibration from one frame and how we evaluate its accuracy on frames that did not contribute to the fit.

\subsection{Camera model}
\label{sec:model}

\begin{figure*}[tbp]
\centering
\includegraphics[width=0.9\textwidth]{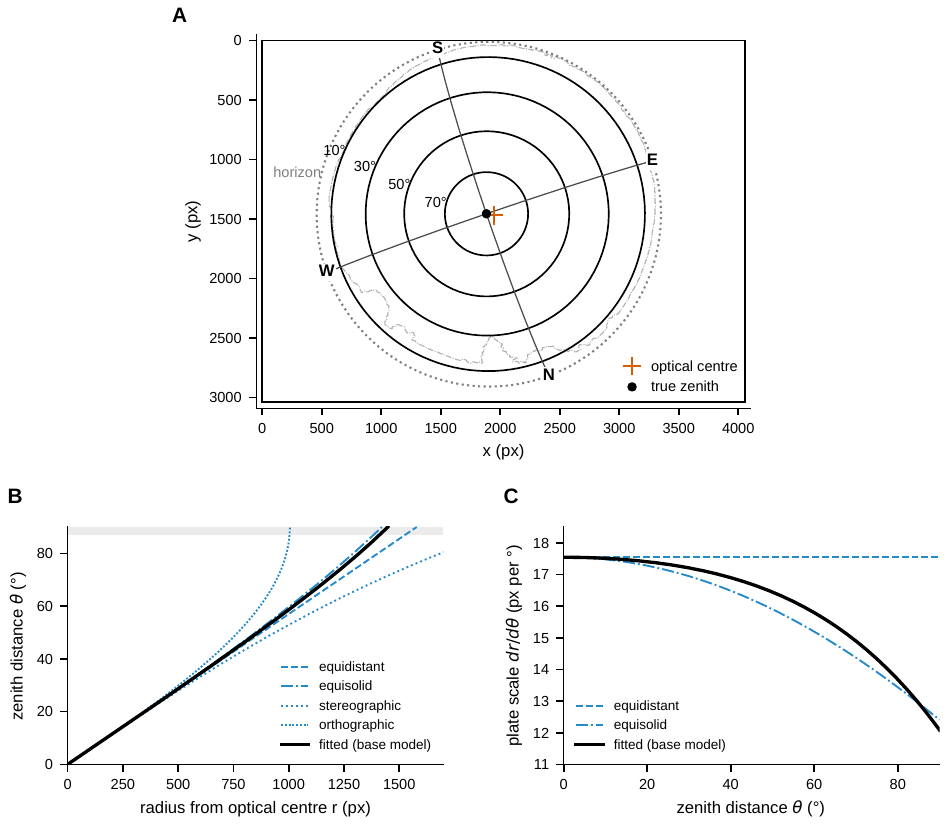}
\caption{Geometry of the base model fitted to the test installation; the offset, tilt and scale shown here are measured properties of this camera, not assumptions of the method. Panel A shows altitude circles and cardinal meridians on the sensor. The black dot marks the true zenith, 65~px from the optical centre (cross) because the optical axis is tilted by 3.7\degree. The grey dotted contour indicates the horizon, and the light-grey dash-dotted contour outlines the valid sky mask. The 10\degree, 30\degree, 50\degree{} and 70\degree{} altitude circles provide reference levels. Panel B compares the camera-frame zenith distance as a function of radius for the fitted lens and four ideal projections with the same focal length. The shaded region, $\theta>87\degree$, corresponds to altitudes below the 3\degree{} matching limit for a levelled camera. With tilt, the boundary also depends on azimuth. Panel C shows how the local radial plate scale decreases across the field, from 17.5~px per degree on the optical axis to 12.1~px per degree at $\theta=90\degree$. This scale converts small angular separations in the radial direction into pixel separations.}
\label{fig:geometry}
\end{figure*}

The model projects a sky direction onto the sensor in four steps (see Fig.~\ref{fig:geometry}). Let $\vect{s}(\altt,\az) = (\cos\altt\,\sin\az,\; \cos\altt\,\cos\az,\; \sin\altt)$ be the unit vector of a star in the topocentric frame ($x$ east, $y$ north, $z$ zenith). We define an eight-parameter model referred to as \emph{base model}, and a ten-parameter model that adds the decentering term, referred to as \emph{extended model}.

\paragraph{Tilt of the optical axis as a rigid rotation}
A camera whose optical axis does not point at the zenith sees the sky rotated. We describe this exactly with a rotation $\Rot \in SO(3)$,
\begin{equation}
\label{eq:rot}
\vect{s}_c = \Rot\,\vect{s}, \qquad \Rot = \Rot_y(\tau_y)\,\Rot_x(\tau_x),
\end{equation}
where $\Rot_x$ and $\Rot_y$ describe rotations about the east--west and north--south axes. To specify the signs used in the implementation,
\begin{equation}
\begin{split}
\Rot_x(t)&=\begin{pmatrix}1&0&0\\0&\cos t&\sin t\\0&-\sin t&\cos t\end{pmatrix},\\
\Rot_y(t)&=\begin{pmatrix}\cos t&0&\sin t\\0&1&0\\-\sin t&0&\cos t\end{pmatrix}.
\end{split}
\end{equation} Two parameters are sufficient because the third degree of freedom, the rotation about the optical axis, is the image rotation $\psi$ introduced below. From $\vect{s}_c$ we obtain the camera-frame altitude and azimuth $(\altt_c, \az_c)$ and the zenith distance $\theta = 90\degree - \altt_c$, the angle between the ray and the optical axis; the total tilt is $\arccos(\cos\tau_x\cos\tau_y)$. This is the formulation of all-sky astrometry used by fireball networks since \citet{ceplecha1987} and \citet{borovicka1995}; here it supplies the orientation component of the model. To recover sky directions from camera-frame directions, we use the transpose matrix, $\Rot^{\mathsf{T}}$.

In Section~\ref{sec:tilt}, we compare the exact rotation with two simplified treatments of tilt: ignoring the tilt and letting the image of the zenith float as a free pixel, or correcting the altitude of each star to first order according to its azimuth while leaving the azimuth unchanged,
\begin{equation}
\label{eq:firstorder}
\altt' = \altt - \tau_x\cos\az - \tau_y\sin\az, \qquad \az' = \az .
\end{equation}
The altitude-only approximation omits the first-order change of azimuth that a tilt also produces. With the conventions of Eq.~(\ref{eq:rot}), away from the zenith,
\begin{equation}
\delta\az = (\tau_y\cos\az-\tau_x\sin\az)\tan\altt + O(\tau^2),
\end{equation}
where the angles are in radians. On the sensor this azimuth change is a tangential displacement of magnitude $r\,|\delta\az|$, which vanishes towards the horizon. Towards the zenith the azimuth coordinate itself becomes singular ($\tan\altt$ diverges) while $r$ tends to zero, so the singularity belongs to the coordinate and not to the displacement, which remains finite. Near the optical axis the altitude and azimuth changes combine into an almost uniform shift of the image by approximately $f\tau$, the leading term of the distance $r(\tau)$ on the sensor between the optical centre and the image of the true zenith; a fit can account for this shift by moving the optical centre. Farther from the axis the displacement changes direction and magnitude with position, so no single shift of the centre reproduces it, and the residuals of the approximate treatments grow towards the horizon. The exact rotation contains both angular changes at all orders, has no singular direction at the zenith, and uses the same two tilt parameters as the altitude-only approximation.

\paragraph{Radial function}
We initially model the lens as rotationally symmetric about the optical axis. We use the generic model of \citet{kannala2006},
\begin{equation}
\label{eq:kb}
r(\theta) = f\left(\theta + k_3\theta^3 + k_5\theta^5\right),
\end{equation}
a truncated odd power series in $\theta$ (in radians) used to approximate the radial projection. Here, $f$ is the focal length in pixels per radian (the plate scale on the axis), and only odd powers appear because $r(\theta)$ must be odd and regular on the axis. The equidistant projection is obtained with $k_3=k_5=0$. The leading cubic coefficients in the Taylor expansions of the equisolid-angle, stereographic and orthographic projections are $-1/24$, $+1/12$ and $-1/6$, respectively (Fig.~\ref{fig:geometry}B). Two coefficients were found sufficient for this lens (a third, $k_7$, changed the residual of the exploratory analysis by about 0.01~px), and Section~\ref{sec:tilt} shows that a degree-5 polynomial in $r$ with three more parameters does not improve the results significantly. We calculate the projected radius directly from Eq.~(\ref{eq:kb}), without an iterative solver. This makes projection efficient when processing thousands of catalogue positions per frame. To recover the angle $\theta(r)$ from a measured radius, we use a few Newton iterations, provided the fitted polynomial is monotonic over the calibrated field, a condition to check for each fitted lens. The derivative $dr/d\theta = f(1 + 3k_3\theta^2 + 5k_5\theta^4)$ is the local radial plate scale (Fig.~\ref{fig:geometry}C) and enters the solid angle seen by each pixel,
\begin{equation}
\label{eq:solid}
\frac{d\Omega}{dA} = \frac{\cos\altt_c}{r}\left|\frac{d\theta}{dr}\right| \quad [\mathrm{sr\,px^{-2}}],
\end{equation}
which is needed to weight cloud fraction and zonal sky brightness. Equation~(\ref{eq:solid}) applies to the rotationally symmetric projection; with the decentering term introduced below, the Jacobian of that correction must be included. At the axis its limiting value is $1/f^2$.

\paragraph{Image rotation and sensor}
The camera-frame azimuth is converted to a direction on the sensor with the rotation $\psi$ of the image with respect to north, $u_0 = r\sin(\psi - \az_c)$, $v_0 = -r\cos(\psi - \az_c)$, with $x$ to the right and $y$ downward. The pixel coordinates are $x = c_x + u$, $y = c_y + v$, where $(c_x, c_y)$ is the optical centre (the image of the optical axis, not of the zenith). In the base model $(u, v) = (u_0, v_0)$; the extended model adds the correction of Eq.~(\ref{eq:bc}) below. The eight parameters $(c_x, c_y, f, \psi, \tau_x, \tau_y, k_3, k_5)$ define the base model.

\paragraph{Optional decentering (extended model)}
\citet{conrady1919} showed that an optical system whose elements are not centred on a common axis produces, besides the radial distortion, a decentering distortion equivalent to a thin prism; \citet{brown1966} gave it the form used in photogrammetry \citep{fraser1997} and in OpenCV \citep{zhang2000,bradski2000opencv}. With normalised coordinates $(\bar u, \bar v) = (u_0, v_0)/f$ and $\bar r^2 = \bar u^2 + \bar v^2$,
\begin{equation}
\label{eq:bc}
\begin{aligned}
\Delta\bar u &= p_1(\bar r^2 + 2\bar u^2) + 2p_2\bar u\bar v, \\
\Delta\bar v &= 2p_1\bar u\bar v + p_2(\bar r^2 + 2\bar v^2),
\end{aligned}
\end{equation}
and $(u, v) = (u_0, v_0) + f(\Delta\bar u, \Delta\bar v)$. The displacement is quadratic in radius and gives the same Cartesian vector at diametrically opposite image positions; its maximum radial amplitude is three times its maximum tangential amplitude. A residual field with these symmetries motivates testing the two additional parameters, although it does not uniquely identify the physical source of the distortion. To recover a sky direction from a pixel position, we first remove the decentering correction in Eq.~(\ref{eq:bc}) using fixed-point iteration. This recovers the coordinates of the rotationally symmetric projection and converges in fewer than eight steps for the fitted coefficients.

\paragraph{Frame of reference}
All parameters refer to the full sensor frame of $4056 \times 3040$ pixels with the origin at the top-left corner and pixel centres at integer coordinates.

\subsection{Star detection}
\label{sec:detection}

We convert each JPEG frame to luminance using the Rec.~601 weights, $0.299R + 0.587G + 0.114B$ \citep{itu2011bt601}. We estimate and subtract the background with a two-dimensional median in boxes of 128~px (\texttt{Background2D} in photutils; \citealt{photutils2024,astropy2022}). We detect stars with DAOStarFinder \citep{stetson1987}, using a Gaussian kernel of 4~px full width at half maximum, a threshold of $4\sigma$ above the background noise, sharpness between 0.2 and 1.0 and roundness within $\pm 0.7$. The last two cuts remove hot pixels and elongated artefacts. The software sets the kernel from the stars of each frame (1.3 times their measured full width, and never less than 4~px). For this camera, whose stars measure 2.8~px, that rule gives the 4~px kernel used throughout. The adaptations needed for other cameras are described in Section~\ref{sec:zeroshot}.

Saturated stars are detected and kept, as their centroids remain accurate. On an example frame, Vega, Capella, Deneb and Fomalhaut, with saturated cores, are detected within 0.5~px of their predicted positions and retained by the fit. In the multi-frame analysis, however, the photometric filter of Section~\ref{sec:association} removes most stars brighter than magnitude~1, because saturation makes them appear fainter than their catalogue magnitude. None of the 31 occurrences of such stars above 10\degree{} of altitude passed it, against 36 of 44 at magnitudes 1.5--2.0. Partly saturated stars that pass the filter have smaller residuals than unsaturated stars of the same magnitude (0.46 against 0.70~px at magnitudes 1--2), so the loss concerns a few dozen bright matches and does not bias the statistics reported below.

The \emph{illuminated lens disc} is the approximately circular region of the sensor onto which the fisheye lens forms an image, bounded by the dark area outside its image circle. It can include foreground obstructions and is not identical to the unobstructed sky. For calibration from a single frame, the software detects this disc by thresholding a blurred, downsampled copy of the image and restricts star detection to that disc. This disc-detection step is included in the calibration package and requires no supplied mask. For the analysis that combines several frames, hereafter the \emph{multi-frame analysis}, we instead supply a custom sky mask. It excludes the dark border, trees and buildings and retains 46\% of the frame area (Fig.~\ref{fig:geometry}A). The calibration package does not create this custom mask. The distinction follows the intended use: one-shot initialisation assumes no additional station-specific input, whereas a dedicated mask is a reasonable preparation step for precise multi-frame calibration. Unlike the automatic disc, it excludes obstructed regions where detections and catalogue associations are less reliable. Most detections are noise peaks or faint stars that play no role after matching. Detection of a full frame takes about 1.5~s on a desktop computer and under a minute on the Raspberry~Pi~4 when the frame is processed in four tiles to bound memory.

\subsection{Reference positions}
\label{sec:catalogue}

We identify stars using 8789 Hipparcos catalogue entries down to magnitude 6.5 \citep{esa1997hipparcos}. We use the distributed positions in ICRS at the catalogue reference epoch J1991.25, which is close to the mean observing epoch of the mission. This epoch specifies when the stellar positions apply. It differs from the orientation of the coordinate axes, for which we approximate ICRS by the J2000 equatorial frame in the precession calculation below. The resulting offsets in 2026 were checked against proper-motion-propagated positions for the 107 named stars of the PyEphem database: they are within about 0.1~px (20~arcsec) except for Arcturus (0.4~px), Sirius and Procyon (0.2~px), and Rigil Kentaurus, which is not observable from the site. A catalogue with positions propagated to the epoch of observation, such as the Gaia DR3 compilation of bright stars \citep{gaia_dr3_best}, would remove these offsets, and could be adopted in future work. We precess the coordinates from J2000 to the date of observation with the IAU~1976 expressions \citep{lieske1977}. The accumulated precession in 2026 is 0.31\degree, about 5~px, so it cannot be neglected. We include the contribution of nutation to sidereal time through the equation of the equinoxes, but do not apply a full nutation correction to the stellar coordinates. We neglect aberration (below 20~arcsec, approximately 0.1~px). We compute the local apparent sidereal time at mid-exposure, and the topocentric altitude and azimuth follow from spherical trigonometry. 
Atmospheric refraction raises a star by about 5~arcmin at 10\degree{} of altitude \citep{bennett1982}, or 1--2~px on this camera. The software compares the detections with apparent altitudes, computed with the formula of \citet{saemundsson1986} scaled to the mean pressure at the elevation of the site. The multi-frame analysis of Sections~\ref{sec:results} and \ref{sec:moon} uses geometric altitudes, so that its numbers remain comparable across the model variants, and Section~\ref{sec:refraction} reports the effect of refraction on the same test matches; the single-frame results of Sections~\ref{sec:single}, \ref{sec:moonzeroshot} and \ref{sec:cameras} use apparent altitudes.

\subsection{Matching of stars and detections}
\label{sec:association}

Matching catalogue stars to detections requires an approximate camera model. We transform the catalogue coordinates to altitude and azimuth using the observing site coordinates and observation time (Section~\ref{sec:catalogue}), then project these directions onto the sensor. The initial model comes from a previous calibration or, for a new camera, from the single-frame procedure in Section~\ref{sec:zeroshot}. We search for detections within a specified radius of each predicted position and require a unique match: the detection must be the nearest candidate to that star, and the star must be the nearest predicted counterpart to that detection. We also apply a photometric filter to reject detections whose measured brightness is inconsistent with the catalogue star, as described next.

For every frame, we first relate the measured detection flux to catalogue magnitude using bright reference stars. We select stars above 30\degree{} of altitude and brighter than magnitude~4.5 with a detection within 12~px of the approximate prediction. We fit the relation $m_\mathrm{inst}=-2.5\log_{10}F=a+m+k(X-1)$, where $F$ is the DAOStarFinder flux, $m$ is catalogue magnitude, $a$ is the photometric zero point, $k$ describes atmospheric extinction and $X$ is the air mass of \citet{kasten1989}. A $2.5\sigma$ clipping step limits the influence of inconsistent reference pairs. The fitted dispersion is 0.4--0.5~mag and the extinction coefficient ranges from 0.3 to 0.9~mag per air mass depending on the night. We then reject a candidate detection if its instrumental magnitude differs from the fitted prediction by more than 0.8~mag. Below 15\degree{} of altitude, we allow a difference of 1.2~mag to account for more variable extinction.

Matching proceeds in two passes. In pass~1, we search for counterparts to stars brighter than magnitude~4 within a radius that grows from 15~px above 45\degree{} of altitude to 80~px at 8\degree, with a stricter photometric filter ($|\Delta m| \le 0.6$), and a pair is accepted only when there is a single candidate and star and detection are mutually nearest. Bright detections are sparse, so a wide radius produces few ambiguities; the same is not true of faint stars, for which a wide radius admitted many noise peaks at low altitude in a first attempt. We fit a seed model robustly to the pairs accepted in pass~1. These magnitude, altitude and radius thresholds are practical choices for this data set rather than physical limits of the method. They select bright, relatively isolated stars and allow larger initial position errors near the horizon. We have not established that they are optimal for other cameras. In pass~2, we use the seed model to search within a fixed radius of 10~px around every catalogue star down to magnitude~5.5 and above 3\degree{} of altitude inside the sky mask, again requiring a single candidate with consistent brightness that is also the nearest star to that detection. These pass-2 pairs form the candidate set for parameter estimation.

Some matches are still wrong or blended, mostly among faint stars and near the horizon. After matching, we first fit the model with a robust loss to limit the influence of these pairs. We then remove training pairs with unusually large residuals within each altitude band and refit the model, following Section~\ref{sec:fit}. This clipping acts only on the fitting data. We retain all accepted matches on the test frame when evaluating predictions, including those with large residuals. The magnitude limit of the matching is therefore a control on matching errors rather than a property of the model. Results are reported for magnitudes of 4.0, 4.5, 5.0 and 5.5.

\subsection{Parameter estimation}
\label{sec:fit}

We estimate the parameters of both the base and extended models from the matched star positions. Both models use the same fitting procedure, which reduces the influence of incorrect matches before the final least-squares fit.

With $N$ pairs $(\altt_i, \az_i) \leftrightarrow (x_i, y_i)$ the parameter vector $\vect{p}$ minimises the residual in pixel space,
\begin{equation}
\label{eq:cost}
S(\vect{p}) = C^2\sum_{i=1}^{N}\sum_{j\in\{x,y\}}\varrho\!\left[\left(\frac{e_{ij}(\vect{p})}{C}\right)^2\right],
\end{equation}
where $\vect{e}_i=\Pi_{\vect{p}}(\altt_i,\az_i)-(x_i,y_i)$ is the predicted-minus-detected pixel displacement, $\varrho$ is applied separately to each coordinate, and $C$ is the robust-loss scale. The residual reported for evaluation and clipping is $\|\vect{e}_i\|$. Fitting in pixels treats the measured centroid coordinates directly; no individual centroid-uncertainty weights are used. We use the trust-region reflective least-squares solver of SciPy \citep{branch1999,scipy2020} with parameter scaling (10~px for the centre and $f$, 0.1\degree{} for $\psi$ and the tilt, 1 for the radial coefficients, $10^{-3}$ for the decentering). We reduce the influence of incorrect matches in three steps: (i) a first fit with the soft-$L_1$ loss $\varrho(z) = 2(\sqrt{1+z}-1)$ and a scale of 3~px, which behaves as least squares for small residuals and as $L_1$ for large ones, which limits the influence of incorrect pairs on the solution; (ii) iterative clipping by altitude band (3--10\degree, 10--20\degree, 20--30\degree, 30--50\degree, 50--70\degree, 70--90\degree): we calculate the median residual $m_b$ and the median absolute deviation $\mathrm{MAD}_b=\mathrm{median}(|d_i-m_b|)$ from the currently retained pairs in each band, where $d_i=\|\vect{e}_i\|$. We retain candidate pairs with residuals below $T_b=\max(3.0\px,\,m_b+3.5\times1.4826\,\mathrm{MAD}_b)$ and refit the model, for up to four rounds. The factor 1.4826 converts the median absolute deviation to a robust estimate of the standard deviation for a Gaussian distribution. Thus 3~px is a minimum threshold, not an upper limit; using a separate threshold in each altitude band accounts for the larger scatter near the horizon and avoids rejecting low-altitude stars simply because their residuals exceed those typical of high-altitude stars; (iii) a final least-squares fit on the retained pairs (about 91\% of the candidates). A fit of the base model on 4000 pairs takes about a second to run.

\subsection{Calibration from a single frame}
\label{sec:zeroshot}

We use the base model for single-frame calibration because this workflow aims to obtain a useful initial solution with minimal input. For this lens, the improvement from the extended model is about 0.1~px, comparable in scale to the difference between single-frame initialisation and the more elaborate multi-frame workflow. The latter also changes the number of images, the mask and the matching filters, so the present experiments do not isolate their individual contributions. We reserve the additional decentering parameters for refinement when several frames support a reproducible improvement.

For a new camera with a visible sky disc, or a field that fills the sensor, and a roughly upward-pointing axis, the software estimates the pose from the image, the site coordinates and the date and time, with no prior calibration. The angular tolerances below were set on the camera of this paper. All pixel radii scale with the plate scale of the frame, so that the same procedure applies to the cameras of Section~\ref{sec:cameras}.
\begin{enumerate}
\item \emph{Detection and sky disc.} Stars are detected as in Section~\ref{sec:detection}, inside the illuminated disc of the lens. The disc gives a first optical centre and, assuming an equidistant lens, a first focal length $f_0 = 2R_h/\pi$ from its radius $R_h$; $k_3 = -0.03$, $k_5 = 0$ and zero tilt complete the initial model. Every pixel radius, search step and robust-loss scale of the procedure is multiplied by $S = f_0/(1005~\mathrm{px\,rad^{-1}})$, so that it keeps the angular value used for this camera ($S\approx1$). When the stars are undersampled (measured full width below 2.2~px), the frame is smoothed with a Gaussian before detection, because DAOStarFinder rejects single-pixel stars as hot pixels; the detection threshold is set from the median absolute deviation of the background-subtracted frame, which, unlike the noise estimate of each 128-px box, is not inflated by gradients from the Moon or from vignetting; and frames above 16~Mpx are detected in $3\times3$ overlapping tiles processed in parallel. FITS files and camera raw files (linear green channel) are read directly.
\item \emph{Orientation search without a prior calibration.} We search a grid of possible camera orientations and scales for the configuration whose predicted bright-star positions best agree with the detections. The image rotation $\psi$ is scanned in steps of 1.5\degree, the focal scale from 0.70 to 1.35 times $f_0$ in steps of 0.03, and the displacement of the zenith from the disc centre up to $\pm0.22\,f_0$ (about 12\degree, since a tilted camera moves the zenith). A pose is scored by the number of catalogue stars brighter than magnitude~3 above 45\degree{} of altitude that fall within about 1.2\degree{} of one of the 250 brightest detections, read from a precomputed distance map of the detections. The grid is searched for both image parities (direct and mirrored, as FITS files and some camera software store the sky). The two best poses of each parity are refined with the radial function fixed and ranked by the number of stars brighter than magnitude~4.5 above 15\degree{} matched within 0.7\degree.
\item \emph{Progressive matching and fit.} Three rounds of matching by position (magnitude $\le 4.5$ within $25S$~px above 15\degree; $\le 5.5$ within $12S$~px above 5\degree; $\le 5.5$ within $7S$~px above 3\degree), each followed by a robust fit of all eight parameters, and the per-band clipping of Section~\ref{sec:fit}. This single-frame workflow uses positional matching throughout; the photometric filter of Section~\ref{sec:association} belongs to the multi-frame workflow.
\item \emph{Quality gate.} A calibration is accepted when its median residual is at most 1.2~px, at least 30 pairs are retained, and either those pairs are at least 15\% of the catalogue stars expected in the field down to the limiting magnitude of the frame (the 90th percentile of the magnitudes of its matched stars), or the selected pose scores at least twice as high as any other pose. With very strong evidence (at least 100 pairs and a pose three times better than any other) the median limit is relaxed to 2~px. Expressing coverage relative to the depth of the frame keeps the gate valid both for a 5-s frame under a full Moon and for a deep frame with a thousand stars.
\item \emph{Hypotheses.} If the gate is not passed, the procedure tries, in order of decreasing prior likelihood, alternative sky discs (a circle circumscribed to the sensor, for lenses whose image circle overfills or is cropped by the sensor; a circle inscribed in it; two circles found by a Hough transform), alternative radial priors for poses that clearly beat their rivals (stereographic-like, $k_3 = +0.04$, and equidistant, $k_3 = 0$), and finally detection with kernels 1.5 and 2 times wider. It stops at the first hypothesis that passes the gate. Two rules skip unpromising branches: if neither the detected nor the circumscribed disc yields a pose clearly better than its rivals, the frame is detected again; and a pose already refined is not refined again under another disc. The search runs within a configurable time budget, after which the calibration is reported as failed with the reason.
\end{enumerate}
For the nine clear frames of this paper the first hypothesis is accepted. Several frames can be calibrated together: the frame with most detections goes through the procedure above, and the model is then refined on all frames jointly.

The blind search is feasible because the sky disc fixes the optical centre and the scale approximately, which leaves an image rotation, a small zenith displacement and a scale factor to be found. That space is small enough to be searched exhaustively by counting bright-star coincidences on one frame, whereas the quadrilateral hashing of Astrometry.net \citep{lang2010astrometry} presupposes a projection close to a tangent plane and the all-sky tools cited in Section~\ref{sec:intro} start from an approximate calibration, user-supplied parameters or long star tracks. The procedure combines a grid search over a small number of orientation and scale parameters with successive fits that include progressively fainter stars.

\subsection{Validation protocol}
\label{sec:protocol}

We evaluate the multi-frame calibration with leave-one-frame-out cross-validation. Each of the nine clear frames is evaluated using a model fitted to the other eight. This measures prediction for an additional frame of the same installation. We repeat the following steps nine times, choosing a different test frame each time:
\begin{enumerate}
\item We fit the pass-1 seed model using bright-star pairs from the eight training frames.
\item We run pass~2 with this seed, separately on the training frames and on the test frame.
\item We apply the clipping described in Section~\ref{sec:fit} only to the training pairs and fit both models to the retained pairs.
\item We evaluate every accepted pass-2 match on the test frame, without removing matches on the basis of their residuals.
\end{enumerate}
We calculate the median, 90th percentile, root mean square and fraction below one pixel from all test-match residuals in the nine evaluations (Table~\ref{tab:bands}). Each frame contributes once, and each accepted match has equal weight; consequently, frames with more matches contribute more to these overall statistics.

We estimate confidence intervals using 2000 bootstrap samples of frames. Each sample contains nine frames selected at random with replacement from the nine evaluated frames. All test residuals belonging to a selected frame are included together; if a frame is selected twice, its residuals appear twice. We calculate the overall median for each sample and use the 2.5th and 97.5th percentiles as the interval limits. This preserves dependence among matches within a frame. Frames from the same night can still share stars and observing conditions, so these intervals describe sensitivity to the sampled frames and do not establish independence between nights. For parameter uncertainties, we use jackknife standard errors from the nine fits, each omitting one frame \citep{efron1982jackknife}. For a parameter $p$, with values $p_{(-i)}$ and mean $\bar p$, the standard error is $[((n-1)/n)\sum_i (p_{(-i)} - \bar p)^2]^{1/2}$ with $n=9$. It measures sensitivity to the sampled frames, rather than absolute calibration uncertainty.

We use the same nine training-and-test splits when comparing magnitude limits, alternative tilt treatments and a degree-5 radial polynomial (Sections~\ref{sec:tilt} and \ref{sec:maglimit}). For single-frame calibration, we fit each frame independently and evaluate the resulting model on the previously selected test matches from the other eight frames, excluding only the calibration frame. We calculate one evaluation median per calibration, then summarise those nine medians (Section~\ref{sec:single}). Finally, we fit models on all nine clear frames for subsequent use and apply them, without refitting, to the separate moonlit and cloudy frames (Section~\ref{sec:moon}).

Two qualifications apply. The evaluation includes only stars that pass detection, the sky mask and the matching filters, so the residuals describe accepted matches rather than all catalogue stars (Section~\ref{sec:checks} quantifies the recovered fraction). In addition, the multi-frame analysis uses a coarse station calibration to identify the photometric reference stars and construct the pass-1 pairs. We reuse this preliminary calibration and these pairs in all nine training-and-test splits, rather than deriving them independently from each set of eight training frames. Section~\ref{sec:checks} examines how these shared inputs affect the selection of candidate matches.

\subsection{Software}
\label{sec:software}

The procedure is released as an open-source Python package, \ascal{} \citep{ascal_repo}, independent of the Lumaria infrastructure. It takes an all-sky image (JPEG, PNG, TIFF, FITS or camera raw), the site coordinates and observation time, and returns the model, the residuals per altitude band and diagnostic figures, through a command-line tool, a Python API and a drag-and-drop web page; it bundles the Hipparcos subset, example frames of the station and an executed notebook. The options that a user may need to change are documented with the symptom that calls for each of them: the time budget, the sky disc (centre and radius), the image parity, the detection kernel and threshold, the tiling of large frames and the refraction correction; the report states which hypothesis of the single-frame procedure was accepted. The catalogue transformations (Julian date, sidereal time, precession) are implemented in the package; the apparent sidereal time agrees with PyEphem \citep{rhodes2011pyephem} to better than 0.1~arcsec on five test dates between 2020 and 2030. The analysis of this paper (Python~3.12 with NumPy, SciPy, Astropy, photutils and OpenCV; timings on an AMD Ryzen~5 5625U laptop and a Raspberry~Pi~4) and its frozen inputs are described in Section~\ref{sec:data_availability}.

\section{Results on clear, moonless frames}
\label{sec:results}

\subsection{Fitted parameters}
\label{sec:parameters}

Table~\ref{tab:parameters} lists the parameters of the base and extended models fitted on all nine frames, and Fig.~\ref{fig:geometry} shows the resulting geometry. The focal length, $f = 1005.2 \pm 0.2$~px per radian, corresponds to 1.558~mm for the 1.55~$\mu$m pixel pitch, within 0.5\% of the manufacturer's nominal value of 1.55~mm. The coefficient $k_3 = -0.021$ places the lens between the equidistant and the equisolid-angle projections (Fig.~\ref{fig:geometry}B), as is common for commercial 180\degree{} fisheye lenses; the radial plate scale is 17.5~px per degree (3.4~arcmin per pixel) on the axis and 12.1~px per degree (5.0~arcmin per pixel) at a camera-frame zenith distance of $90\degree$, which is reached at $r = 1448$~px. The optical axis is tilted 3.73\degree{} from the zenith, $(\tau_x, \tau_y) = (-1.815\degree, -3.255\degree)$, which places the true zenith 65~px from the optical centre; the image is rotated by $\psi = 161.055\degree$ with respect to north. The decentering coefficients of the extended model are only about two to three jackknife standard errors from zero and correspond to a displacement at the edge of the field of 1.7~px in the radial direction and 0.6~px in the tangential direction; the extended model changes the other parameters by less than 0.5~px or 0.02\degree.

\begin{table}
\centering
\caption{Parameters of the base and extended models fitted to all nine clear frames, with jackknife standard errors (s.e.) from the nine fits that each omit one frame. Centre coordinates are in pixels, $f$ is in pixels per radian, and rotation and tilt angles are in degrees. The radial and decentering coefficients are dimensionless, with the latter expressed in units of $10^{-4}$ for coordinates normalised by $f$. The horizontal line below $p_2$ separates the fitted parameters from derived quantities: the physical focal length, total tilt, zenith position and image radius at $\theta=90\degree$.}
\label{tab:parameters}
\footnotesize
\setlength{\tabcolsep}{4pt}
\begin{tabular}{lrrrr}
\toprule
 & \multicolumn{2}{c}{Base model} & \multicolumn{2}{c}{Extended model} \\
Parameter & value & s.e. & value & s.e. \\
\midrule
$c_x$ (px) & 1948.26 & 0.19 & 1948.64 & 0.26 \\
$c_y$ (px) & 1467.98 & 0.13 & 1468.337 & 0.078 \\
$f$ (px rad$^{-1}$) & 1005.24 & 0.22 & 1005.16 & 0.29 \\
$\psi$ (\degree{}) & 161.05445 & 0.00085 & 161.05356 & 0.00074 \\
$\tau_x$ (\degree{}) & $-1.8146$ & 0.0050 & $-1.8252$ & 0.0044 \\
$\tau_y$ (\degree{}) & $-3.2549$ & 0.0084 & $-3.2612$ & 0.0092 \\
$k_3$ & $-0.02098$ & 0.00067 & $-0.02065$ & 0.00095 \\
$k_5$ & $-0.00512$ & 0.00040 & $-0.00539$ & 0.00061 \\
$p_1$ ($\times 10^{-4}$) & -- & -- & $-2.0$ & 1.1 \\
$p_2$ ($\times 10^{-4}$) & -- & -- & $-1.85$ & 0.57 \\
\midrule
Focal length (mm) & \multicolumn{2}{c}{1.558} & \multicolumn{2}{c}{1.558} \\
Total tilt (\degree{}) & \multicolumn{2}{c}{3.73} & \multicolumn{2}{c}{3.74} \\
Zenith position (px) & \multicolumn{2}{c}{(1884, 1456)} & \multicolumn{2}{c}{(1884, 1457)} \\
$r(90\degree{})$ (px) & \multicolumn{2}{c}{1448} & \multicolumn{2}{c}{1447} \\
\bottomrule
\end{tabular}
\end{table}

\subsection{Accuracy on test frames}
\label{sec:accuracy}

\begin{figure*}[tbp]
\centering
\includegraphics[width=0.85\textwidth]{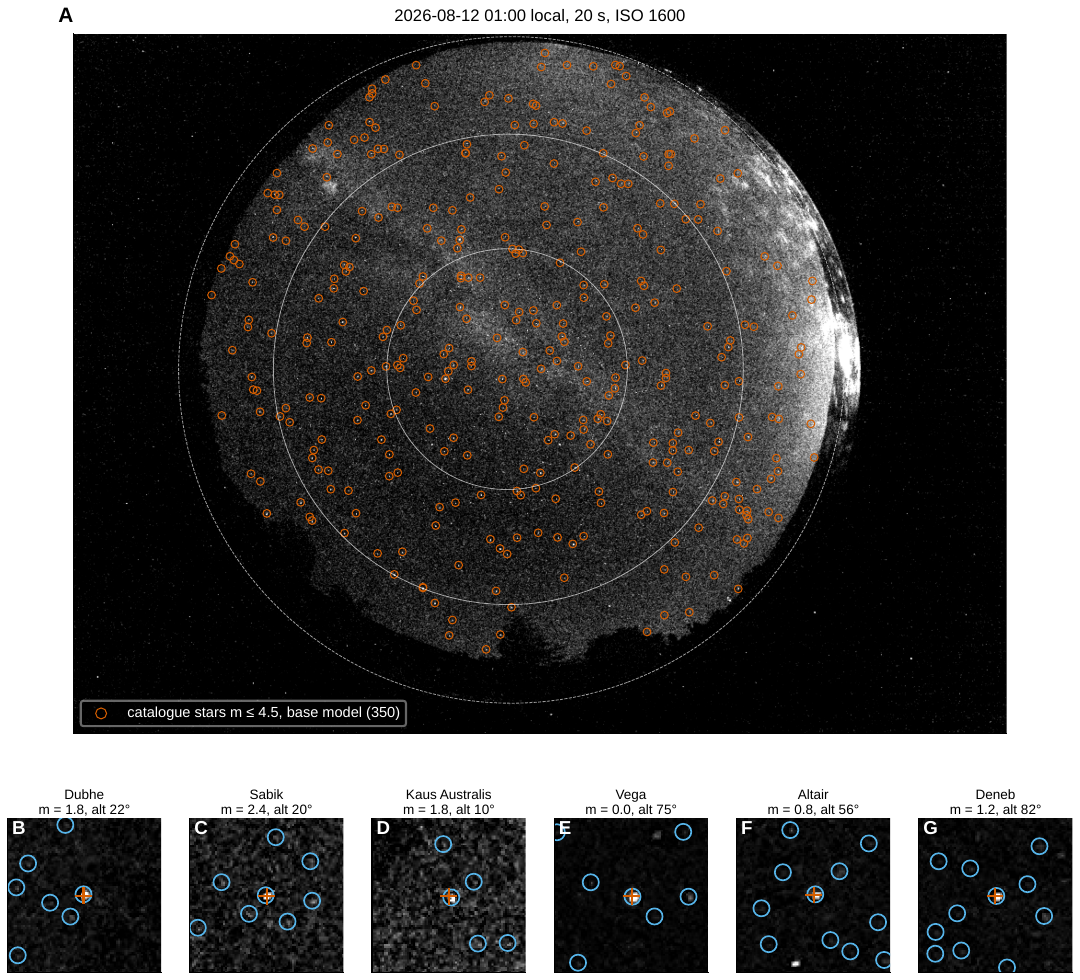}
\caption{Example frame taken on 12 August 2026 at 01:00 local time. Panel A shows the all-sky camera field. Orange circles mark the positions predicted by the base model for 350 catalogue stars brighter than magnitude 4.5. The altitude circles indicate 0\degree, 30\degree{} and 60\degree. The bright arc on the right is the dome illuminated by a nearby lamp. Panels B--G show $80\times80$~px regions around six bright stars at low (B--D) and high (E--G) altitude. Blue circles indicate detections from the algorithm, and orange crosses mark the predicted positions of the reference catalogue stars.}
\label{fig:example}
\end{figure*}

Fig.~\ref{fig:example} shows one of the frames with the catalogue stars projected by the base model and, in the cut-outs, the detections around six bright stars at low and high altitude together with the predicted positions. We find that the agreement is within a pixel from the zenith down to 10\degree{} of altitude.

\begin{figure*}[tbp]
\centering
\includegraphics[width=0.85\textwidth]{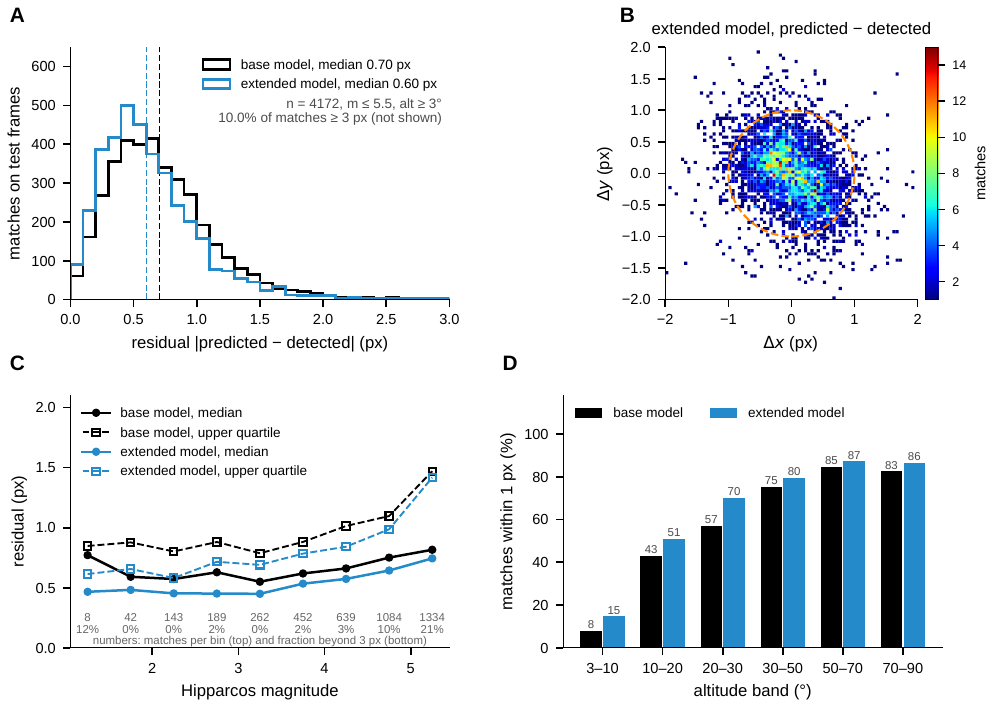}
\caption{Residuals for 4172 matches from nine frames, with each frame evaluated using models fitted on the other eight. Panel A shows the distribution of residual magnitudes. The dashed black and blue lines indicate the medians for the base and extended models, respectively. The 10\% of matches with residuals above 3~px lie outside the plotted range. Panel B shows the two residual components for the extended model. The dashed circle has a radius of 1~px and contains 77\% of the matches. Panel C shows the median and upper quartile of the residuals in pixels as a function of stellar magnitude, together with the number of matched objects per bin and the fraction with residuals above 3~px. Panel D shows the fraction of matched stars with residuals below one pixel in each altitude band, for the base model in black and the extended model in blue.}
\label{fig:residuals}
\end{figure*}

Fig.~\ref{fig:residuals} and Table~\ref{tab:bands} summarise the prediction accuracy on test frames. The base model has a median residual of $0.70\px$ (95\% frame-bootstrap interval 0.64--0.80~px) on the 4172 test matches to magnitude 5.5, with 72\% of them within one pixel; for stars brighter than magnitude~4 the median is $0.59\px$ and 84\% are within one pixel. The extended model reaches $0.60\px$ (0.53--0.71~px), with 77\% within one pixel, and $0.48\px$ for the bright stars. Its improvement in the overall median is $0.10\px$ (0.09--0.11~px), and it improves the median in every test frame. At the plate scale of the axis, $0.70\px$ corresponds to 2.4~arcmin.

\begin{table*}[tbp]
\centering
\caption{Prediction residuals by altitude band for stars down to magnitude 5.5. Each frame is evaluated with models fitted on the other eight frames. For both models, we report the median and 90th percentile (p90) in pixels and the proportion of matched stars with residuals below 1~px. The final column summarises the single-frame calibrations (apparent altitudes): for each altitude band, we first calculate the median residual on the other frames for each calibration, then report the median of those nine values (Section~\ref{sec:single}).}
\label{tab:bands}
\begin{tabular}{lrrrrrrrr}
\toprule
Altitude & $n$ & \multicolumn{3}{c}{Base model} & \multicolumn{3}{c}{Extended model} & Single frame \\
band (\degree) & & median & p90 & $<1$~px & median & p90 & $<1$~px & median \\
\midrule
3--10 & 75 & 2.60 & 9.10 & 8\% & 2.44 & 9.85 & 15\% & 1.76 \\
10--20 & 412 & 1.18 & 8.48 & 43\% & 0.97 & 8.36 & 51\% & 1.23 \\
20--30 & 641 & 0.90 & 6.28 & 57\% & 0.71 & 6.57 & 70\% & 0.95 \\
30--50 & 1323 & 0.69 & 1.55 & 75\% & 0.61 & 1.47 & 80\% & 0.75 \\
50--70 & 1198 & 0.57 & 1.12 & 85\% & 0.52 & 1.10 & 87\% & 0.64 \\
70--90 & 523 & 0.64 & 1.15 & 83\% & 0.52 & 1.06 & 86\% & 0.72 \\
\midrule
All & 4172 & 0.70 & 3.04 & 72\% & 0.60 & 2.96 & 77\% & 0.77 \\
\bottomrule
\end{tabular}
\end{table*}

The residual grows towards the horizon (Table~\ref{tab:bands}): about 0.6--0.7~px above 30\degree, 1~px between 10 and 30\degree{} and 2.6~px below 10\degree, a band that holds fewer than 2\% of the matches. The low-altitude bands have long residual tails, with 90th percentiles of 6--10~px. Two observations suggest that detection and matching contribute to these tails: only 1.5\% of matches to stars brighter than magnitude~4 exceed 3~px, whereas the fraction is much larger when fainter stars are included (Section~\ref{sec:maglimit}); and adding decentering reduces the typical residual but leaves similar large-residual tails. Faint detections near the sky-mask boundary can have uncertain centroids, overlap another source, or be associated with the wrong catalogue star. These mechanisms could produce errors that neither geometric model removes. This interpretation is suggestive, not a measurement of the number of incorrect matches or evidence that all large residuals are non-geometric. The lowest band is also where the plate scale is smallest, the extinction and refraction gradients largest, and the stars pass behind the dome edge, trees and the light domes of nearby villages, so geometric error cannot be separated from these effects with the present residuals.

\subsection{Treatment of the tilt}
\label{sec:tilt}

\begin{figure*}[tbp]
\centering
\includegraphics[width=0.9\textwidth]{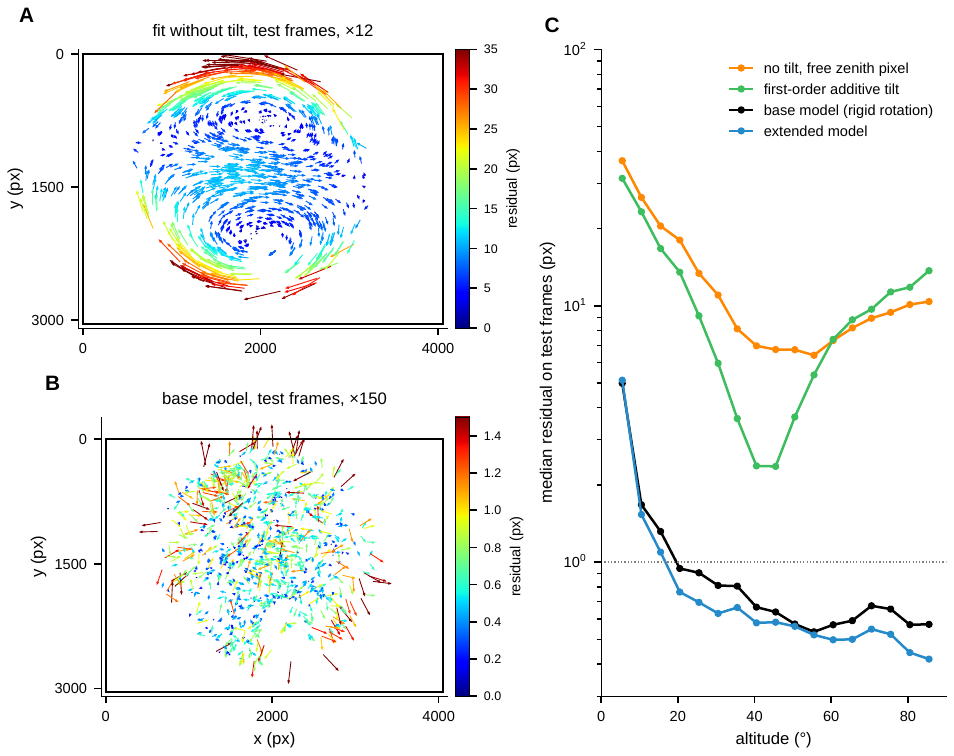}
\caption{Residuals for different treatments of the all-sky camera tilt, evaluated on frames excluded from fitting. Panel A shows residual vectors for a model that ignores tilt and allows the zenith position to vary freely, using stars brighter than magnitude~4. We magnify the arrows by a factor of 12, so a 10~px residual appears as a 120~px arrow. Panel B shows the base-model residual vectors for the same stars with residuals below 2~px (98\% of the sample), magnified by a factor of 150. Panel C compares the median residual by altitude for the three tilt treatments and the extended model. The dotted horizontal line marks a residual of one pixel.}
\label{fig:tilt}
\end{figure*}

Our camera's optical axis is tilted by 3.7\degree{} from the zenith. To assess how the calibration must account for this misalignment, we fit three model variants to the same training data and evaluate them on the same test matches (Table~\ref{tab:structure}, Fig.~\ref{fig:tilt}). The first variant is the base model with $\tau_x=\tau_y=0$ fixed, leaving six free parameters: $(c_x,c_y,f,\psi,k_3,k_5)$. With zero tilt, the optical centre coincides with the projected zenith, whose pixel coordinates remain free. This restricted version of the same model gives a median residual of 8.8~px overall and 16~px below 30\degree{} altitude. The second variant applies a first-order correction to altitude while leaving azimuth unchanged. Despite having eight parameters, it gives median residuals of 6.5~px overall and 12~px below 30\degree. Its fitted tilt falls to 0.8\degree{} and its centre shifts by about 47~px to compensate for the incomplete correction. Our base model instead represents the tilt as an exact rotation of sky directions. With the same eight parameters, it reduces the median residual to $0.70\px$ overall and $1.0\px$ below 30\degree. Thus the 6--9~px residuals describe the two simplified model variants tested on this installation, rather than a general limit for calibration methods that omit tilt.

The residual field of the no-tilt fit (Fig.~\ref{fig:tilt}A) has the structure predicted in Section~\ref{sec:model}, a swirl around the zenith and a displacement of the outer ring, with two points of small residual symmetric about the centre where the unabsorbed part of the rotation field changes sign; the rotation removes it (Fig.~\ref{fig:tilt}B). A degree-5 polynomial in $r$ with the exact rotation gives $0.69\px$ with three additional parameters, so the radial function is not the limiting element. The experiment demonstrates calibration without precise levelling for this installation; it does not compare separate levelled and tilted cameras.

\begin{table*}[tbp]
\centering
\caption{Comparison of model structures using the same 4172 test matches. We fit each model on eight frames and evaluate its predictions on the remaining frame, repeating this for all nine frames. Residual statistics combine these nine evaluations. KB denotes the Kannala--Brandt radial function in Eq.~(\ref{eq:kb}).}
\label{tab:structure}
\begin{tabular}{lcccccc}
\toprule
 & Free & \multicolumn{3}{c}{Median residual (px)} & rms & Within \\
Model structure & param. & 3--30\degree & 30--90\degree & all & (px) & 1~px \\
\midrule
KB, zenith pixel only (no tilt) & 6 & 16.12 & 7.92 & 8.76 & 12.54 & 1\% \\
KB, first-order additive tilt & 8 & 12.38 & 4.98 & 6.48 & 10.29 & 4\% \\
KB, rigid rotation (base model) & 8 & 1.03 & 0.64 & 0.70 & 2.48 & 72\% \\
Degree-5 polynomial in $r$, rigid rotation & 11 & 1.02 & 0.62 & 0.69 & 2.48 & 71\% \\
KB, rigid rotation + decentering (extended model) & 10 & 0.80 & 0.55 & 0.60 & 2.48 & 77\% \\
\bottomrule
\end{tabular}
\end{table*}

\subsection{Magnitude limit}
\label{sec:maglimit}

The magnitude limit plays two roles, as shown in Table~\ref{tab:maglimit} and Fig.~\ref{fig:robustness}A. Restricting the \emph{training} set to brighter stars does not change the fitted geometry: the median residual for the full test population moves by 0.02~px when the limit is tightened from 5.5 to 4.0. Evaluating only brighter stars, with the training limit set to the same magnitude as in Table~\ref{tab:maglimit}, reduces the contribution of ambiguous matches and noisy centroids: the fraction of matches beyond 3~px falls from 10\% at magnitude 5.5 to 1.5\% at magnitude 4, and the median from 0.70 to $0.61\px$ (base model, 85\% within one pixel). When using predicted star positions to measure sky transparency or cloud coverage, a limit of magnitude 4--4.5 keeps 120--200 stars per frame with 1--2\% of large residuals, and is the setting we use in practice.

\begin{table*}[tbp]
\centering
\caption{Effect of the magnitude limit used to select fitting stars. For each limit, we fit on eight frames and evaluate predictions on the remaining frame, repeating this for all nine frames. We report the median residual in pixels and the proportion of matched stars with residuals below 1~px, both for all test stars down to magnitude 5.5 and for those brighter than the fitting limit. For the latter group, we also give the proportion with residuals above 3~px for the base model.}
\label{tab:maglimit}
\begin{tabular}{lrrrrrrrrr}
\toprule
 & \multicolumn{4}{c}{Test stars $m \le 5.5$ ($n = 4172$)} & \multicolumn{5}{c}{Test stars $m \le$ cut} \\
 & \multicolumn{2}{c}{Base} & \multicolumn{2}{c}{Extended} & & \multicolumn{3}{c}{Base} & Extended \\
Cut & median & $<1$~px & median & $<1$~px & $n$ & median & $<1$~px & $>3$~px & median \\
\midrule
$m \le 4.0$ & 0.72 & 71\% & 0.61 & 77\% & 1102 & 0.61 & 85\% & 1.5\% & 0.49 \\
$m \le 4.5$ & 0.72 & 71\% & 0.60 & 77\% & 1762 & 0.63 & 81\% & 1.9\% & 0.52 \\
$m \le 5.0$ & 0.72 & 71\% & 0.60 & 77\% & 2852 & 0.67 & 76\% & 5.0\% & 0.56 \\
$m \le 5.5$ & 0.70 & 72\% & 0.60 & 77\% & 4172 & 0.70 & 72\% & 10.0\% & 0.60 \\
\bottomrule
\end{tabular}
\end{table*}

\subsection{Atmospheric refraction}
\label{sec:refraction}

The multi-frame results above use geometric altitudes, so the fitted radial function absorbs part of the refraction. That absorption is incomplete, because refraction is symmetric about the local zenith whereas the radial function is centred on the tilted optical axis. Table~\ref{tab:refraction} repeats the leave-one-frame-out evaluation on the same test matches with apparent altitudes \citep{saemundsson1986}. The overall median changes little (0.707 to $0.701\px$ for the base model in this re-run, whose baseline differs from Table~\ref{tab:bands} by 0.005~px because it re-implements the folds), but the low-altitude bands improve: the 3--10\degree{} median falls from 2.55 to 2.12~px for the base model and from 2.40 to 2.05~px for the extended model. Refraction is therefore a measurable part of the low-altitude residual of the multi-frame analysis, and the software applies it by default.

\begin{table*}[tbp]
\centering
\caption{Effect of atmospheric refraction on the prediction residuals, with the same nine training-and-test splits and test matches. Apparent altitudes follow \citet{saemundsson1986} for a site elevation of 650~m. The medians are expressed in pixels by altitude band. The last column is the fraction of matches within 1~px.}
\label{tab:refraction}
\begin{tabular}{lcrrrrrrr}
\toprule
Model & Refraction & 3--10\degree & 10--20\degree & 20--30\degree & 30--50\degree & 50--90\degree & All & $<1$~px \\
\midrule
Base & no & 2.55 & 1.19 & 0.89 & 0.69 & 0.60 & 0.707 & 71.4\% \\
Base & yes & 2.12 & 1.14 & 0.88 & 0.68 & 0.59 & 0.701 & 72.2\% \\
Extended & no & 2.40 & 0.95 & 0.70 & 0.60 & 0.52 & 0.604 & 77.3\% \\
Extended & yes & 2.05 & 0.88 & 0.69 & 0.61 & 0.52 & 0.604 & 77.5\% \\
\bottomrule
\end{tabular}
\end{table*}

\subsection{Decentering and frame-to-frame stability}
\label{sec:stability}

\begin{figure*}[tbp]
\centering
\includegraphics[width=0.8\textwidth]{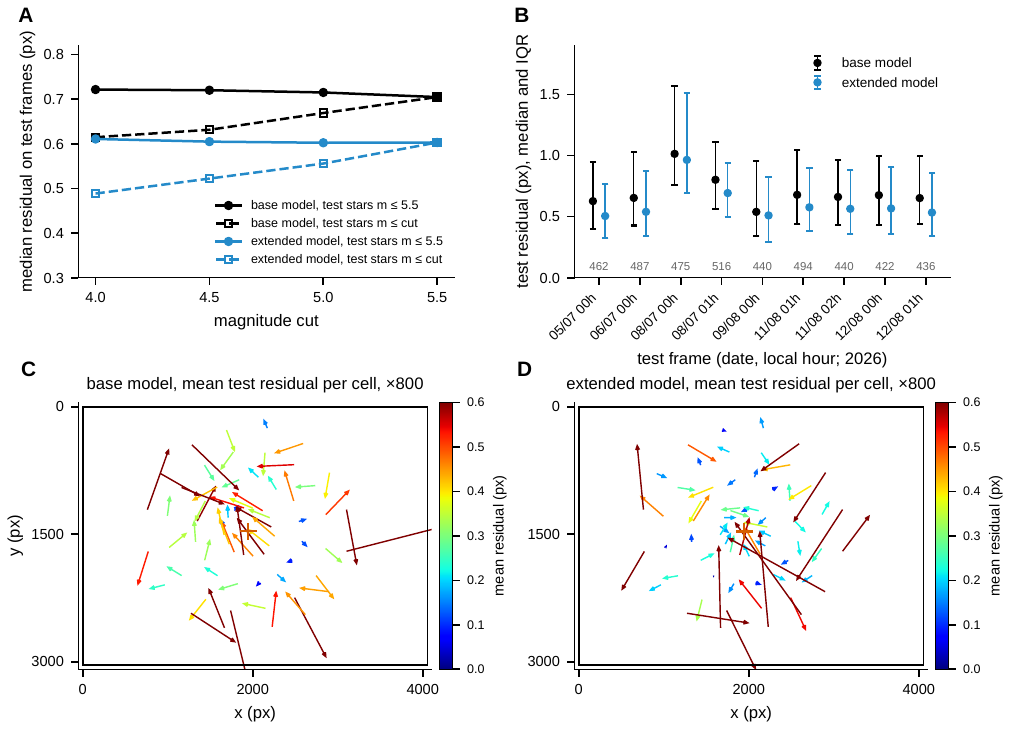}
\caption{Effects of magnitude limit, test frame and decentering correction on predictions for frames excluded from fitting. Panel A shows the median residual as a function of the training magnitude limit. Solid lines with circles include all test stars, whereas dashed lines with squares include only test stars brighter than that limit. Panel B shows the median and interquartile range for each test frame, using models fitted on the other eight frames. Panels C and D show the mean residual vector per sensor cell for the base and extended models, respectively. Cells combine four radial bins with outer edges at 450, 800, 1100 and 1480~px and 16 azimuth sectors. We show cells with at least 20 matches and magnify vectors by a factor of 800, so a 0.5~px mean residual appears as a 400~px arrow. The base model leaves the decentering pattern described by Eq.~(\ref{eq:bc}).}
\label{fig:robustness}
\end{figure*}

Fig.~\ref{fig:robustness}C shows why the extended model is offered as an option: the mean residual of the base model per cell of the sensor is a coherent field of 0.3--0.6~px pointing in the same direction at opposite azimuths and growing with radius, the signature of Eq.~(\ref{eq:bc}), and the extended model removes most of that residual field (Fig.~\ref{fig:robustness}D). The gain of $0.1\px$ in the median is real but small, and with nine frames the decentering coefficients are only about two to three jackknife standard errors from zero. We recommend the base model as the default and the extended model when the residual map shows this pattern and the improvement is reproducible across frames.

The per-frame medians of the base model range from 0.54 to 1.01~px and those of the extended model from 0.50 to 0.96~px (Fig.~\ref{fig:robustness}B). The largest median occurs in the 8 July frame at 00:00 local time for both models. Section~\ref{sec:moon} tests the nine-frame models on the frames of the same period that were excluded from the clear-sky set.

\subsection{Calibration from a single frame}
\label{sec:single}

\begin{figure*}[!t]
\centering
\includegraphics[width=\textwidth]{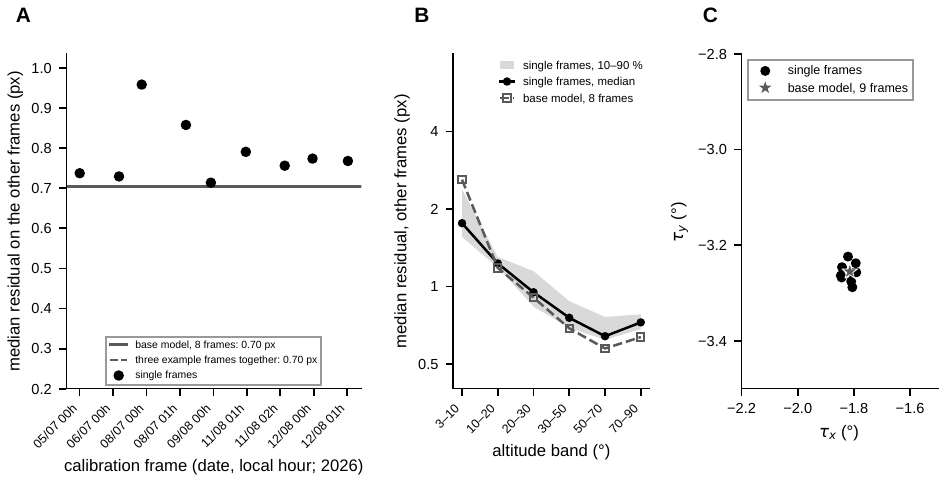}
\caption{Calibration from a single frame. Panel A shows the median residual on the other frames for each independently calibrated frame. The solid line gives the overall prediction residual from fitting the base model to eight frames and evaluating it on the ninth. The dashed line gives the result from fitting the three example frames together and evaluating the eight unused clear frames. Panel B compares residuals by altitude band, showing the median and 10th--90th percentile across the single-frame calibrations alongside the eight-frame fits. Panel C compares the tilt angles $(\tau_x,\tau_y)$ recovered from individual frames with the final base-model fit to all nine frames. Thus panels A and B use eight-frame fits to measure prediction accuracy, whereas panel C uses the nine-frame fit as a parameter reference. The standard deviations across the nine single-frame fits are 0.021\degree{} in $\tau_x$ and 0.018\degree{} in $\tau_y$, corresponding to approximately 0.35~px at the on-axis plate scale. Single-frame calibrations use apparent altitudes.}
\label{fig:zeroshot}
\end{figure*}

We calibrated each of the nine frames independently, with no prior model, supplied mask or labels, and evaluated the resulting calibration on accepted test matches from the other eight frames (Fig.~\ref{fig:zeroshot}, Table~\ref{tab:single}). These calibrations use apparent altitudes. The evaluation matches are converted to apparent altitudes in the same way. All nine pass the quality gate at the first hypothesis, with about 500 pairs and a median residual of 0.51--0.59~px on their own frame. For each calibration, we calculate one median over all evaluation matches in the other eight frames. These nine evaluation medians range from 0.71 to 0.96~px, and their median is $0.77\px$; all nine are below one pixel. In the lowest band, 3--10\degree, the median of the nine calibrations is 1.76~px (Table~\ref{tab:bands}). The multi-frame base model gives an overall test-match median of $0.70\px$ when fitted to eight frames. These summaries use different weighting: one value per calibration for single-frame results, and one value per accepted test match for multi-frame results. Table~\ref{tab:bands} gives the corresponding altitude-dependent results. The parameters recovered from independent frames agree to 0.4~px in the optical centre, 0.2~px in $f$, 0.003\degree{} in $\psi$ and 0.02\degree{} in the tilt (Fig.~\ref{fig:zeroshot}C).

The three example frames provided with the software (8 July, 28 July and 9 August; Section~\ref{sec:clearset}) can also be fitted jointly with the same automatic procedure. Only the 8 July frame belongs to the nine-frame clear-sky set; the other two belong to the moonlit set. We therefore exclude that one clear-sky frame and evaluate the joint calibration on the eight unused clear-sky frames (3656 test matches). The fit retains 1294 pairs with a median fitting residual of 0.59~px, and its evaluation median is $0.70\px$, with 71\% of matches within one pixel. This is an illustrative combination of three frames, not a systematic test of how accuracy varies with the number of calibration frames.

\begin{table*}[tbp]
\centering
\caption{Single-frame calibrations of the nine clear frames, scored on the test matches of the other frames.}
\label{tab:single}
\footnotesize
\begin{tabular}{ll}
\toprule
Quantity & Result \\
\midrule
Frames calibrated alone & 9 (9 accepted; 9 below 1~px on the other eight frames) \\
Pairs per fit (median, range) & 506 (469--513) \\
Fit residual on the frame itself (median) & 0.54~px \\
Evaluation median over nine calibrations (IQR; range) & 0.77~px (0.74--0.79; 0.71--0.96) \\
Same, stars brighter than magnitude 4 & 0.67~px \\
Matches within 1~px on the other eight frames (median) & 67\% \\
Spread of parameters (s.d. over 9 fits): $c_x$, $c_y$, $f$ & 0.3, 0.4, 0.2~px \\
Spread of parameters (s.d. over 9 fits): $\psi$, $\tau_x$, $\tau_y$ & 0.003\degree, 0.021\degree, 0.018\degree \\
\midrule \multicolumn{2}{l}{Three example frames fitted jointly (one clear-sky frame and two moonlit frames)} \\
Retained fitting pairs; median fit residual & 1294; 0.59~px \\
Evaluation on the eight unused clear-sky frames & 0.70~px (71\% within 1~px) \\
\bottomrule
\end{tabular}
\end{table*}

\subsection{Interpretation of the residuals}
\label{sec:checks}

Two checks qualify the accuracy figures above. The first concerns the population they describe: the residuals are computed on accepted matches, not on all catalogue stars, so it matters how many of the stars in the field are actually matched. Table~\ref{tab:recovery} counts, for the nine frames, the catalogue stars to magnitude 5.5 above 3\degree{} inside the sky mask according to the base model and those that received a pass-2 match: 41\% overall, 62\% of the stars brighter than magnitude~4, rising from 28\% of the bright stars in the 3--10\degree{} band to 89\% above 70\degree. The unmatched stars are mostly undetected on a 20~s, 8-bit frame or fail the uniqueness and photometric filters. The residuals therefore characterise the sky above 30\degree{} well for bright stars and the horizon sparsely, and the fraction of matches with residuals above 3~px reported in Section~\ref{sec:maglimit} and Table~\ref{tab:maglimit} is a property of the matched population, not a measured false-match rate.

\begin{table*}[t]
\centering
\caption{Catalogue stars expected in each of the nine frames (magnitude $\le 5.5$, altitude $\ge 3\degree$, inside the sky mask), counted once per frame, and those that received a pass-2 match (matched/expected), by altitude band and magnitude. The 4172 matched candidates correspond to 1174 distinct stars.}
\label{tab:recovery}
\footnotesize
\begin{tabular}{lrrrr}
\toprule
Altitude band (\degree) & $m \le 4$ & $4 < m \le 5$ & $5 < m \le 5.5$ & All to 5.5 \\
\midrule
3--10 & 39/140 (28\%) & 20/305 (7\%) & 16/304 (5\%) & 75/749 (10\%) \\
10--20 & 144/302 (48\%) & 169/678 (25\%) & 99/640 (15\%) & 412/1620 (25\%) \\
20--30 & 162/297 (55\%) & 284/724 (39\%) & 195/713 (27\%) & 641/1734 (37\%) \\
30--50 & 335/517 (65\%) & 532/1168 (46\%) & 456/1305 (35\%) & 1323/2990 (44\%) \\
50--70 & 281/368 (76\%) & 528/921 (57\%) & 389/915 (43\%) & 1198/2204 (54\%) \\
70--90 & 141/158 (89\%) & 217/352 (62\%) & 165/347 (48\%) & 523/857 (61\%) \\
All & 1102/1782 (62\%) & 1750/4148 (42\%) & 1320/4224 (31\%) & 4172/10154 (41\%) \\
\bottomrule
\end{tabular}
\end{table*}

The second check concerns the coarse calibration that the multi-frame matching uses to find its photometric anchors, and the pass-1 pairs derived from it, which are shared by all folds. Replacing that coarse calibration, frame by frame, with the frame's own single-frame model (which uses no prior calibration) changes the zero points by 0.01~mag and leaves pass-2 match sets with a Jaccard similarity of 0.97, and re-running the pass-1 rule with it reproduces 96\% of the pass-1 pairs. The candidate matches are therefore relatively insensitive to the source of the coarse calibration in this comparison. The complete preprocessing and fitting procedure has not been repeated independently within each fold.

\section{Results for frames with the Moon and with clouds}
\label{sec:moon}

Most frames of an unattended station are not taken under the conditions of Section~\ref{sec:results}. The 73 frames of Section~\ref{sec:moonset}, with moonlight or clouds, are used 1) to test whether the calibration obtained from clear, moonless frames predicts the stars that remain visible when the Moon is up or clouds are present, and 2) to test whether the single-frame procedure can obtain a calibration from such frames. In both cases the reference is the clear-sky set. The two sets contain different frames but share some nights; the pass-2 matches of these frames are scored with the nine-frame base model, and each single-frame calibration is evaluated on the test matches of all nine clear frames. No calibration frame is present in that reference set; sharing an acquisition night does not remove a frame from evaluation.

\subsection{Transfer of the clear-sky calibration}
\label{sec:transfer}

Each of the 73 frames was processed as a frame of the multi-frame analysis (detection inside the sky mask, photometric anchors and pass-2 matching with the nine-frame base model as seed, no geometric refit), and the residuals of both models were measured (Table~\ref{tab:transfer}, Fig.~\ref{fig:moon}A). The geometry calibrated on clear frames predicts the stars visible on moonlit frames with residuals comparable to those on the clear frames. The 32 frames of the clear class, 25 of them with the Moon between 2\degree{} and 37\degree{} of altitude at phases from 18\% to full, give an overall median of $0.62\px$ with 77\% of the matches within one pixel, independently of the altitude of the Moon. What the Moon changes is the number of matched stars, about 350 per frame against 460 without it. Frames with low cloud coverage show similar residuals. Partial cloud coverage increases the median residual moderately ($0.82\px$) while approximately halving the number of matches; the bright stars keep a median of 0.5~px in all three classes. With heavy and overcast skies the matching still returns 25--115 pairs per frame, but with medians of 2--7~px. These pairs are likely to include many chance coincidences with noise peaks or cloud features that fall within the search radius of a catalogue star, and the photometric filter does not remove them because, with few real stars, the zero point fitted on the frame is itself unreliable. An unusable frame is therefore recognised by the median residual of its matches, not by their number alone, and an analysis that reuses a calibration should apply such a check.

\begin{table*}[tbp]
\centering
	\caption{Clear-sky calibration applied to the 73 moonlit and cloudy frames through the pass-2 matching, by cloud class. Median residual values are indicated in pixels. The column labelled ``Usable'' shows the number of selected frames with at least 15 photometric anchors and 20 matches.}
\label{tab:transfer}
\begin{tabular}{lrrrrrrrr}
\toprule
Cloud class & Frames & Usable & Matches/frame & \multicolumn{2}{c}{Base model} & \multicolumn{2}{c}{Extended model} & Base, $m \le 4$ \\
 & & & (median) & median & $<1$~px & median & $<1$~px & median \\
\midrule
Clear ($\le 0.1$) & 32 & 32 & 374 & 0.62 & 77\% & 0.52 & 82\% & 0.51 \\
Low cloud coverage ($0.1$--$0.3$) & 14 & 14 & 164 & 0.63 & 74\% & 0.52 & 78\% & 0.51 \\
Partial ($0.3$--$0.6$) & 14 & 14 & 174 & 0.82 & 57\% & 0.74 & 61\% & 0.51 \\
Heavy ($0.6$--$0.9$) & 3 & 3 & 114 & 5.33 & 30\% & 5.17 & 31\% & 0.73 \\
Overcast ($>0.9$) & 10 & 6 & 42 & 6.42 & 12\% & 6.48 & 12\% & 4.60 \\
\bottomrule
\end{tabular}
\end{table*}

\subsection{Calibration from moonlit and cloudy frames}
\label{sec:moonzeroshot}

\begin{figure*}[tbp]
\centering
\includegraphics[width=\textwidth]{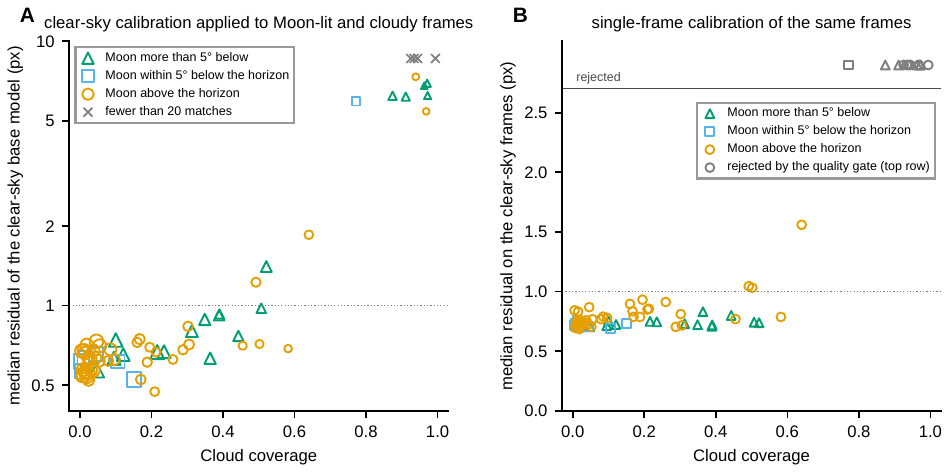}
	\caption{Performance in moonlit and cloudy frames. A: median residual of the clear-sky base model on the matches of each frame against its cloud coverage. Grey crosses mark frames with fewer than 20 accepted matches; they are placed at a fixed vertical position for visibility and excluded from the transfer statistics. Coloured markers indicate the position of the Moon relative to the horizon in both panels; marker size represents the number of matches only in panel A. B: single-frame calibration of the same images, showing the median residual of each accepted calibration on the nine reference frames. Rejected calibrations are shown in grey in the strip at the top. The dotted line in each panel marks a residual of one pixel.}
\label{fig:moon}
\end{figure*}

The single-frame procedure of Section~\ref{sec:zeroshot} was applied unchanged to the 73 frames (Table~\ref{tab:moonzeroshot}, Fig.~\ref{fig:moon}B). It accepts all 32 clear frames, all 14 frames with low cloud coverage, all 14 partially clouded frames and one of the 13 heavy or overcast frames. Every accepted calibration of the clear and low-cloud classes, and 12 of the 14 partially clouded ones, has an evaluation median below 1~px on the nine reference frames. The median across the 61 accepted calibrations is $0.75\px$ and the worst is 1.56~px. The optical centres and tilt angles recovered from images with no or low cloud coverage agree to about 1~px and 0.05\degree, respectively. None of the accepted calibrations has a wrong orientation: the largest evaluation medians (1.03--1.04~px for two partially clouded frames, 1.56~px for the heavily clouded one, with 120 matches) come from coarse but correct poses. The remaining 12 heavy or overcast frames are rejected because no hypothesis passes the gate.

\begin{table*}[tbp]
\centering
	\caption{Single-frame calibration of the 73 moonlit and cloudy frames, by cloud class. The ``Outliers'' column counts calibrations accepted with an evaluation median of at least 1.5~px on the clear frames. Median and worst: the median and maximum, respectively, of the evaluation medians of accepted calibrations on the nine clear frames.}
\label{tab:moonzeroshot}
\begin{tabular}{lrrrrrr}
\toprule
Cloud class & Frames & Accepted & $<1$~px & Outliers & Median (px) & Worst (px) \\
\midrule
Clear (Moon up 25, near 4, down 3) & 32 & 32 (25, 4, 3) & 32 & 0 & 0.74 & 0.87 \\
Low cloud coverage & 14 & 14 & 14 & 0 & 0.79 & 0.93 \\
Partial & 14 & 14 & 12 & 0 & 0.76 & 1.04 \\
Heavy and overcast & 13 & 1 & 0 & 1 & 1.56 & 1.56 \\
All & 73 & 61 & 58 & 1 & 0.75 & 1.56 \\
\bottomrule
\end{tabular}
\end{table*}

\section{Other all-sky cameras}
\label{sec:cameras}

The single-frame procedure was developed on the camera of this paper. To generalise it to other instruments, we collected one clear night frame from each of 23 all-sky systems in public archives and data releases, and calibrated each with the procedure of Section~\ref{sec:zeroshot}. The only inputs are the site coordinates and the time of the exposure, and we used default options.

\subsection{Test set}
\label{sec:camset}

Table~\ref{tab:systems} lists the systems of the set and the results of this generalisation test. The set includes the camera of this paper; five observatory sky monitors: the MMT all-sky camera \citep{mmt_skycam}, the MASCOT and ALPACA cameras of the ESO Paranal Observatory \citep{eso_archive}, the SkyCamA of the Liverpool Telescope \citep{lt_skycam} and KLCAM at Dome~A \citep{yang2025superwide}; four fireball-network cameras: a Desert Fireball Network station, from the data release of the Madura Cave meteorite fall \citep{devillepoix2021madura}, and three FRIPON/SCAMP stations, from that of the Winchcombe fall \citep{devillepoix2022winchcombe}; and 13 stations of the Allsky software network \citep{jacquin_allsky}, with Raspberry~Pi HQ and ZWO cameras and lenses of 1.55--2.8~mm. The frames span 0.3 to 77~Mpx, monochrome and colour sensors, 8- to 16-bit data in JPEG, PNG, FITS and camera-raw formats, focal lengths of 257--2930~px per radian, star widths of 1.4--5.2~px and camera tilts of 0.6--6.5\degree. They include a full Moon (FRIPON), urban skies, overlays printed on the image, laser guide stars (Paranal), the telescope structure in the field (Liverpool), image circles cut by the sensor (DFN, several Allsky stations) and FITS files that store the sky mirrored. Each calibration was checked by applying it, without refitting, to a second frame of the same camera from another hour or night, and that of the DFN station also against the astrometry of the DFN pipeline for the Madura Cave fireball.

\begin{table*}[tbp]
\centering
\caption{All-sky systems of the test set and their single-frame calibration. Stars: matched stars retained by the fit; median: median residual of that fit; other frame: median residual of the same calibration, without refitting, on another frame of the camera (when it gives at least 20 matches).}
\label{tab:systems}
\footnotesize
\setlength{\tabcolsep}{4pt}
\begin{tabular}{llllrrr}
\toprule
System & Camera and lens & Image (px) & Format & Stars & Median (px) & Other frame (px) \\
\midrule
Lumaria ZRO (this work) & RPi HQ, EDATEC 1.55~mm & $4056\times3040$ & JPEG & 528 & 0.53 & 0.87 \\
MMT Observatory & ZWO ASI294MC Pro, 1.8~mm & $1422\times1411$ & FITS & 745 & 0.21 & 0.23 \\
ESO MASCOT, Paranal & monochrome CCD & $580\times512$ & FITS & 361 & 0.12 & 0.12 \\
ESO ALPACA, Paranal & monochrome, 16 bit & $8750\times8750$ & FITS & 453 & 1.21 & -- \\
Liverpool Telescope SkyCamA & Starlight Xpress CCD & $1392\times1040$ & FITS & 652 & 0.26 & 0.27 \\
KLCAM, Dome~A & DSLR & $3476\times5208$ & JPEG & 828 & 0.39 & -- \\
DFN DFNEXT029 & Nikon D810, Samyang 8~mm & $7380\times4928$ & NEF & 919 & 0.45 & 0.93 \\
FRIPON GBWL01, Cardiff & Basler acA1300, 1.25~mm & $1296\times966$ & FITS & 34 & 0.30 & -- \\
FRIPON ENSW01, Honiton & DMK 23G445, 1.25~mm & $1280\times960$ & FITS & 84 & 0.42 & -- \\
FRIPON ENNW01, Manchester & Basler acA1300, 1.25~mm & $1296\times966$ & FITS & 38 & 0.56 & -- \\
Allsky Cosmotori & ZWO ASI178MC, 2.5~mm & $3096\times2080$ & JPEG & 465 & 0.78 & 0.82 \\
Allsky Kamuela & ZWO ASI224MC, 150\degree{} lens & $1304\times976$ & JPEG & 504 & 0.39 & 0.65 \\
Allsky Heiligenfelde & RPi HQ (binned), 1.56~mm & $2028\times1520$ & JPEG & 566 & 0.33 & 0.37 \\
Allsky Triebes & RPi HQ, 2.8~mm & $4056\times3040$ & JPEG & 317 & 0.45 & 1.46 \\
Allsky Great Strickland & ZWO ASI676MC, 2.5~mm & $3552\times3552$ & JPEG & 988 & 0.36 & 0.36 \\
Allsky UZO & ZWO ASI224MC, stock lens & $1304\times976$ & JPEG & 387 & 0.37 & 0.36 \\
Allsky Stalag-13 & RPi HQ (binned), 1.8~mm & $2028\times1520$ & JPEG & 269 & 0.42 & 0.47 \\
Allsky Riphaeus & RPi HQ, 1.55~mm & $4056\times3040$ & JPEG & 331 & 0.62 & 0.67 \\
Allsky Bottenbach & ZWO ASI678MC, fisheye & $3840\times2160$ & PNG & 714 & 1.24 & 1.29 \\
Allsky Castiglione della Pescaia & ZWO ASI676MC, 2.5~mm & $1776\times1776$ & JPEG & 699 & 0.48 & 1.24 \\
Allsky Nideggen & RPi HQ, 1.55~mm & $4056\times3040$ & JPEG & 353 & 0.71 & 0.79 \\
Allsky Konkoly, Budapest & RPi HQ, 1.56~mm & $4056\times3040$ & JPEG & 398 & 0.39 & 0.47 \\
Allsky Frank & RPi HQ, 1.85~mm & $4056\times3040$ & JPEG & 458 & 0.85 & 0.97 \\
\bottomrule
\end{tabular}
\end{table*}

\subsection{Results}
\label{sec:camresults}

\begin{figure*}[tbp]
\centering
\includegraphics[width=\textwidth]{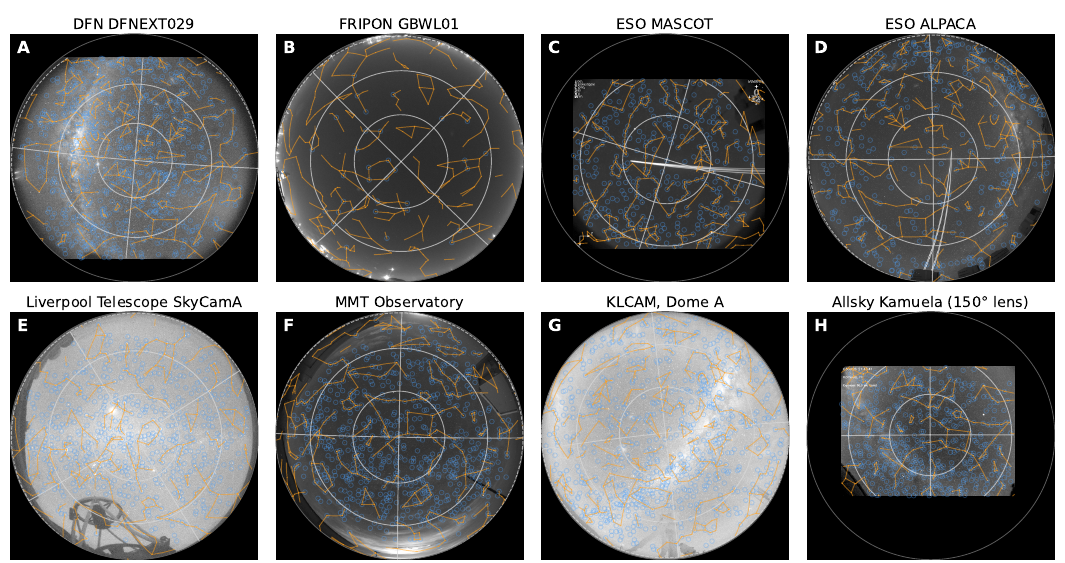}
\caption{Single-frame calibrations of eight systems of Table~\ref{tab:systems}. Blue circles mark the detections matched to catalogue stars; the white lines are the altitude circles at 0\degree{} (dashed), 30\degree{} and 60\degree{} and the north--south and east--west lines, and the orange lines are the constellation figures, all drawn with the fitted model. A: Desert Fireball Network station (Nikon D810 raw frame, image circle cut by the sensor). B: FRIPON/SCAMP video camera under a full Moon (34 matches). C and D: ESO Paranal MASCOT (0.3~Mpx) and ALPACA (77~Mpx), with laser guide stars. E: Liverpool Telescope SkyCamA, with the telescope in the field. F: MMT all-sky camera, a mirrored FITS frame. G: KLCAM at Dome~A. H: an amateur station with a 150\degree{} lens whose field fills the sensor.}
\label{fig:cameras}
\end{figure*}

The procedure calibrates all 23 frames (Table~\ref{tab:systems}, Fig.~\ref{fig:cameras}). The first hypothesis was accepted for 14 frames. The others needed pre-smoothing of undersampled stars (7 frames), the mirrored parity (5), the disc circumscribed to the sensor (4), a second or third detection with a wider kernel (3) or a non-equisolid radial prior (1, the DFN lens). Evaluating every hypothesis exhaustively instead found the same solutions. The median fit residuals range from 0.12~px (MASCOT) to 1.24~px, with a median of 0.42~px, or 0.9--5.6~arcmin. The transfer of a calibration to other frames was tested on the 18 systems for which a second frame of the same camera, from another hour or night, gives at least 20 matches. Applied without refitting, the calibrations predict the stars of that frame with medians of 0.12--1.46~px (median 0.66~px). 
The second frames of the FRIPON cameras have the full Moon high in the sky and give only 8--13 matches, so those three calibrations are supported by their mutual consistency ($f = 339$--346~px per radian for the three identical lenses) rather than by a second frame. The DFN calibration agrees with the DFN pipeline astrometry of the Madura Cave fireball to about 2~arcmin (about 1~px) at 5--20\degree{} of altitude. Two further images from Alcor all-sky cameras, for which no second frame is available, are also calibrated: 684 stars at 1.33~px for an ALPHEA camera whose JPEG frame is contrast-stretched by quadrants, and 795 stars at 0.28~px for a mirrored monochrome frame.

\subsection{Field of view and unsuitable frames}
\label{sec:fov}

The procedure requires a fisheye projection and an optical axis within about 12\degree{} of the zenith, but not a complete image circle. Two stations of the set have lenses whose field fills the sensor: the Kamuela camera, with a lens sold as 150\degree, sees 52\% of the sky above the horizon (44\degree{} from the axis at the edges of the short side of the sensor and 82\degree{} at its corners). It is calibrated with 504 stars at 0.39~px, predicting another frame to 0.65~px. The Triebes camera, a 2.8~mm lens on a Raspberry~Pi HQ sensor, sees 49\% of the sky and is calibrated with 317 stars at 0.45~px (1.46~px on a second frame with only 55 matches). Both lenses are close to equidistant ($|k_3| \le 0.02$). Two rectilinear wide-angle cameras of the same collection, a ZWO ASI178 with a 3.2~mm lens and a Raspberry~Pi Camera Module~3 Wide, could not be calibrated with our procedure: a gnomonic projection, $r = f\tan\theta$, lies outside the radial priors and diverges towards 90\degree, and in a narrower field the tilt becomes harder to separate from a shift of the optical centre, which it resembles near the axis (Section~\ref{sec:model}). Wide-angle cameras of this kind, including phone cameras, would need a different radial function and initialisation and are outside the scope of this work.

Frames of the collection with too few usable stars are rejected when no hypothesis passes the gate: overcast or heavily clouded skies, dew drops or dirt on the dome, strong underexposure, a full Moon in an urban sky with 5-s exposures, and fields mostly blocked by trees or buildings. For cameras with few stars per frame, such as the FRIPON video cameras under a full Moon, calibrating several frames of the night together is the recommended use.

\section{Discussion}
\label{sec:discussion}

\subsection{Contribution and comparison with existing methods}

Exact orientation models and automatic star identification are established elements of all-sky astrometry. Fireball networks have described the optical axis by a rigid rotation combined with a radial polynomial since \citet{borovicka1995}, and their recent descendants refine that scheme with automatic catalogue matching. Some of them are PRISMA \citep{barghini2019}, FRIPON, whose fisheye model uses a ninth-order odd radial polynomial with two non-symmetric terms \citep{jeanne2019fripon}, the Global Fireball Observatory \citep{devillepoix2020gfo} and the Global Meteor Network \citep{vida2021gmn}; site-monitoring and cloud cameras have adopted comparable solutions \citep{antuna2022orion,yin2025ali,yang2025superwide}. The present work combines them in a compact, open workflow for a low-cost monitoring camera and evaluates prediction across frames. The developed eight-parameter base model captures the dominant geometry of this installation, while two optional decentering terms reduce the median residual by $0.10\px$. The tilt comparison shows why an altitude-only approximation, which has the same number of parameters as the exact rotation, or a model that represents the tilt only through the position of the zenith pixel as ORION does, is insufficient for a camera 3.7\degree{} off the zenith: such models are adequate for an instrument levelled by an operator, but not for the tilt observed here. 

Moreover, the method developed here for single-frame initialisation reduces the observations needed to obtain a usable calibration. All nine clear-frame calibrations are accepted, and each predicts the other eight frames with an evaluation median below one pixel. The median of these nine medians is $0.77\px$, compared with an overall test-match median of $0.70\px$ for the multi-frame base model. This modest difference supports the intended division between rapid initialisation and subsequent refinement. A joint fit to the three example images provided with the software gives $0.70\px$ on eight unused clear frames; repeated trials with other combinations would be needed to establish how generally that result holds. The same procedure calibrates, with default options, frames from 23 all-sky systems of very different design (Section~\ref{sec:cameras}), which is the main evidence that it is not specific to the camera on which it was developed.

We also ran the astrometric fit of SkyFit2 \citep{rms_software} on the same star--detection pairs and the same nine training-and-test splits. With a similar number of parameters, its radial models and our base model give equivalent prediction residuals (0.68--0.69~px with 9--10 parameters, against $0.70\px$ with 8, with overlapping intervals), and its best configuration, a seventh-order radial model with asymmetry terms (11 parameters), reaches 0.64~px, against $0.60\px$ for our extended model (10 parameters). The difference between the two approaches therefore lies not in the accuracy of the model, but in the automatic initialisation from one frame without manual identification of stars, the explicit treatment of the tilt and the validation of predictions on frames not used for fitting.

Near the optical axis, one pixel corresponds to approximately 3.4~arcmin (Section~\ref{sec:parameters}). Multiplying the test residuals of 0.70~px for the base model and 0.60~px for the extended model by this local angular scale gives approximately 2.4 and 2.1~arcmin, respectively, using the unrounded values. These conversions provide an angular interpretation of the pixel residuals near the optical axis. They do not measure the angular error over the whole field, because the conversion varies with position and differs between radial and tangential directions. Published results include sub-pixel residuals for PRISMA cameras \citep{barghini2019}, the Ali Observatory camera \citep{yin2025ali} and the KLCAM all-sky camera at Dome~A (0.4~px; \citealt{yang2025superwide}), while the ORION example reports about 9~arcmin, or 1.7~px \citep{antuna2022orion}. Differences in hardware, sky coverage, match selection and validation limit a direct comparison.

\subsection{Performance with Moon and clouds}

Section~\ref{sec:moon} answers two practical questions. First, a calibration obtained from clear, moonless frames can be reused on the moonlit frames that make up most of a station's archive, provided that the camera geometry remains unchanged. The stars that remain visible are predicted to $0.62\px$, independently of the altitude and phase of the Moon. Low cloud coverage mainly reduces the number of matches, whereas partial cloud coverage also degrades the residual distribution. This is expected for a camera that has not moved, since the Moon changes the background and not the geometry, but it is the property that allows a station to select its calibration frames freely and to reuse the resulting model on other frames, as long as enough stars remain visible for reliable matches. Second, a calibration can be obtained directly from every moonlit frame of the sample and from all the partially clouded ones, while overcast frames are rejected. Under the Moon, the alternative disc hypotheses of the procedure compensate for a less reliable automatic estimate of the sky disc. A mask remains useful for the multi-frame analysis: we obtained it from a daytime image of the same camera, in which a simple edge or contrast detection separates the sky from the dome edge, trees and buildings.

\subsection{Limitations and interpretation}

The detailed evaluation concerns one camera and lens at one observing site over six weeks. Applying the cloud-free, moonless selection criteria to our observations yields a small reference data set of nine frames acquired on six dates. This sample supports an assessment of the method on the tested installation, including the comparison of model structures. The 23 systems of Section~\ref{sec:cameras} show that the single-frame procedure generalises, but each of them is represented by one or two frames, so they test the calibration and its transfer to another frame rather than its long-term stability on those cameras. The bootstrap intervals (0.64--0.80~px for the base model) reflect that size. The results support a compact model for this installation, but do not determine the radial order or decentering correction required by other lenses. Jackknife parameter errors measure sensitivity to the sampled frames, but do not include uncertainty in the reference catalogue, timestamps, model choice or shared preprocessing, whose influence is partially evaluated in Section~\ref{sec:checks}, through the overlap of the candidate matches.

The overall residual is dominated by well-detected stars above 30\degree. Below 10\degree{} fewer than 2\% of the matches contribute and the median residual is about 2.6~px. Low-altitude scatter can reflect centroid errors, extinction, refraction, obstruction, blending and model inadequacy, and the residual distribution alone does not identify which matches are wrong. Independent checks of star identification would be needed to separate calibration error from detection and matching performance there. The JPEG tone curve and compression may affect centroid precision, but we have not assessed their contribution through a comparison with raw images.

Correcting for refraction, by fitting apparent altitudes, reduces the 3--10\degree{} residual by about 0.4~px (Section~\ref{sec:refraction}). The refraction correction uses the mean pressure at the site and no temperature, so the fitted coefficients remain an effective mapping under the sampled conditions rather than a laboratory measurement of the lens, particularly near the horizon.

Single-frame initialisation assumes a fisheye projection and an optical axis within about 12\degree{} of the zenith. It does not require a complete image circle, and has been tested down to fields that cover half of the sky (Section~\ref{sec:fov}). Rectilinear wide-angle lenses are outside its scope. The photometric filter of the multi-frame matching needs sufficient bright reference stars, and when it is fitted on few anchors, as on overcast frames, it no longer rejects chance coincidences. The quality gate remains the most delicate element of the single-frame procedure: with fewer than about 40 pairs, a wrong pose can give a tight fit, and the margin of the selected pose over its rivals is what rejects it; the weakest margin among the accepted calibrations of Section~\ref{sec:cameras} is 2.15. Finally, the colour channels of the sensor are combined into one luminance image; calibrating each Bayer channel separately, with colour transformations to the catalogue photometry, would allow transparency to be mapped over the sky, and is left for future work.

\section{Summary and conclusions}
\label{sec:summary}

We present an open workflow for calibrating low-cost all-sky cameras with an eight-parameter fisheye model, automatic star matching and single-frame initialisation. Its distinguishing contribution, relative to the methods reviewed in Section~\ref{sec:discussion}, is the joint demonstration of explicit camera-tilt correction, automatic one-shot calibration without a prior model, and robustness to the tested lunar illumination and partial cloud cover, while retaining sub-pixel median prediction residuals. These capabilities are evaluated together within the same openly available workflow, with quality checks that identify most unusable images.

On nine cloud-free, moonless frames, the base model yields a median residual of $0.70\px$ across 4172 test matches, reduced to $0.60\px$ by two decentering parameters. For the tested 3.7\degree{} camera tilt, the exact rotation substantially improves on both a free zenith pixel and an altitude-only correction. All nine single-frame trials pass the quality gate and predict positions on the other eight frames with evaluation medians below one pixel; the median of these nine values is $0.77\px$. Applying the clear-sky calibration to 73 frames with varying lunar illumination or cloud cover gives median residuals of $0.62\px$ on moonlit, cloud-free images and $0.82\px$ under partial cloud. A calibration can therefore be obtained from clear, moonless observations and reused while the camera geometry remains unchanged and enough stars remain visible for reliable matching. Direct calibration from moonlit or partly cloudy images is also possible: 61 of the 73 frames are calibrated, none with a wrong orientation (Section~\ref{sec:moon}). The same single-frame procedure calibrates, with default options, one frame from each of 23 all-sky systems of very different design, from 0.3 to 77~Mpx, with a median fit residual of $0.42\px$ and a median of $0.66\px$ on a second frame of the same camera; correcting the catalogue positions for refraction improves the residuals near the horizon.

The procedure provides the astrometric calibration for the first Lumaria station and will be applied to subsequent units. For a new station, we recommend initialising from a clear frame and checking the number of matches, the fitted orientation and the residuals by altitude. Once several clear, moonless frames are available, the calibration can be refined with a dedicated sky mask and evaluated by fitting to all but one frame in turn. The exported model should be accompanied by the number of frames and matches, the acquisition dates, the residuals by altitude and the model version. Periodic single-frame checks, particularly after maintenance, can identify changes that require recalibration. For star-visibility measurements, a magnitude limit of 4--4.5 is a useful compromise; search radii should follow the altitude-dependent residual quantiles in Table~\ref{tab:bands}.

The main limitations are the larger residuals near the horizon, the small reference sample of the detailed evaluation, and the fact that each of the other cameras is represented by one or two frames. Future work will evaluate the procedure over long periods on additional Lumaria installations and other networks, check the appropriate model order for each lens, and use the colour channels of the sensor to map sky transparency. We encourage users to test the software on images from other all-sky systems, which are the most effective way to extend the procedure. The implementation is released as \ascal.

\section*{Acknowledgements}

We thank the ZRO team and Luc\'ia's family for hosting the Lumaria station. This work made use of NumPy, SciPy \citep{scipy2020}, Astropy \citep{astropy2022}, photutils \citep{photutils2024}, OpenCV \citep{bradski2000opencv}, PyEphem \citep{rhodes2011pyephem} and Matplotlib.
Most of the Python code used in this analysis was written with assistance from AI models (Anthropic Claude and OpenAI Codex). These tools were also used in preparing the manuscript, including drafting of some of the sections, language and grammar revision and critical review of the text and results. All analysis code was reviewed by the authors. The numerical results are produced by analysis scripts intended for release with the frozen data, and the validation protocol (evaluation on frames excluded from fitting, single-frame calibrations, moonlit and cloudy frames, and frames from other all-sky cameras) was designed to test the results independently of the tools used to obtain them.

\section*{Data availability}
\label{sec:data_availability}

The calibration software, with the example frames, is available at \url{https://github.com/GonzalezFJR/ascal}. The data and scripts of this paper are deposited in Zenodo at \url{https://doi.org/10.5281/zenodo.23764194}: the frozen inputs of the analysis (the frames of the clear-sky and the moonlit and cloudy sets, their detections, bright-star pairs of pass~1, sky mask, Hipparcos subset), the test matches with the residual of each model, the per-frame parameter sets, the frame bootstrap, the model-structure comparison, the single-frame calibrations, the transfer and single-frame results on the 73 moonlit and cloudy frames, and the scripts that reproduce every number, table and figure of this paper, with the software used for each result. The frames of the test set of Section~\ref{sec:cameras} come from public archives and data releases \citep{devillepoix2021madura,devillepoix2022winchcombe,eso_archive,lt_skycam,mmt_skycam} and from public image archives of stations of the Allsky network; the deposit lists their sources and contains the fitted model, the matched stars and diagnostic figures of every system.

\bibliographystyle{aasjournal}
\bibliography{references}

\end{document}